\documentclass[preprint,12pt]{elsarticle}

\usepackage{amssymb}
\usepackage{amsthm}

\usepackage{amsmath}
\usepackage{amsbsy}
\usepackage{mathrsfs}
\usepackage{graphicx}
\usepackage{bm}
\usepackage{subfigure}
\usepackage{color}

\newcommand{\diag}{\mbox{diag}}

\newcommand{\eg}{\textit{e.g.}}

\newcommand{\ustb}{%
Department of Applied Mathematics and Mechanics, University of Science and Technology Beijing,
Beijing 100083, China}

\newcommand{\csrc}{%
Beijing Computational Science Research Center,
Beijing 100193, China}

\newcommand{\csrca}{\csrc}

\newcommand{\camtsinghua}{%
Zhou Pei-Yuan Center for Applied Mathematics,\\
Tsinghua University, Beijing 100084, China}

\journal{Elsevier}

\begin{document}

\begin{frontmatter}

\title{A family of single-node second-order boundary schemes
for the lattice Boltzmann method}


\author{Weifeng Zhao}
\ead{wfzhao@ustb.edu.cn}
\address{{\ustb}}

\author{Wen-An Yong
\corref{cor1}
}
\ead{wayong@tsinghua.edu.cn}
\address{{\camtsinghua};}
\address{{\csrca}}
\cortext[cor1]{Corresponding author}

\begin{abstract}

In this work, we propose a family of single-node second-order boundary schemes for the lattice Boltzmann method with general collision terms.
The construction of the schemes is quite universal and simple, it does not involve concrete lattice Boltzmann models and uses the half-way bounce-back rule as a central step.
The constructed schemes are all second-order accurate if so is the bounce-back rule.
In addition, the proposed schemes have good stability thanks to convex combinations.
The accuracy and stability of several specific schemes are numerically validated for multiple-relaxation-time models in both 2D and 3D.
\end{abstract}


\end{frontmatter}



\section{Introduction}
\label{sec:intro}

The lattice Boltzmann method (LBM) is an efficient technique for modeling complex fluid flows and has attracted much attention in a variety of fields \cite{LKS,YMLS,CD} because of its easy implementation and second-order accuracy \cite{HL1997-1,HL1997-2}.
In using the method, a fundamental problem is how to treat boundary conditions (BCs) since almost each flow occurs in a region with boundaries. Typical examples are the no-slip BCs for particulate flows \cite{Ladd1,Ladd2}, the wetting BCs for two-phase flows \cite{Jacqmin1999,Huang2013} and those for free interface problems \cite{Bogner2015}.
Fortunately, due to its kinetic origin, the LBM can naturally accommodate many different BCs for flows with complicated geometries.
This is a prominent advantage of the LBM over other conventional numerical methods for fluid dynamics.

In the literature, there are various different boundary schemes accompanying the lattice Boltzmann method. The schemes involve either only the current lattice node or other neighboring lattice ones.
The latter does not obviously apply to the situation where no enough neighboring nodes are available, as pointed out in \cite{Junk2005pre}.
The former is referred to as single-node boundary schemes.
The widely used one is the bounce-back rule proposed in \cite{Ladd1, Ladd2}. This scheme usually has first-order accuracy unless the boundary locates at the middle of two neighboring nodes.
Other single-node boundary schemes can be found in \cite{Junk2005pre,Noble1995,Inamuro1995,Ginzeburg1996,GSPK2015,ZY2017}.
Those in \cite{Noble1995,Inamuro1995,Ginzeburg1996} are of second-order accuracy but only for straight boundaries, while that in \cite{Junk2005pre} uses the DFs of all directions and needs to compute, at each boundary node, the inverse of a matrix with entries given by complicated formulas. In our recent work \cite{ZY2017}, we constructed a class of single-node boundary schemes with second-order accuracy
for curved boundaries by using the Maxwell iteration \cite{Yong2016pre} for the two-relaxation-time (TRT) model \cite{IGinzburg2005,IGinzburg2008_1,IGinzburg2008_2}.
The constructions of the boundary schemes in \cite{Ginzeburg1996,Junk2005pre,ZY2017} rely heavily on the Chapman-Enskog expansion, asymptotic analysis or the Maxwell iteration.
On the other hand, in \cite{GSPK2015} a different construction was proposed by combining interpolations and the half-way bounce-back rule (the boundary locates at the middle of two neighboring nodes).
We remark that the construction in \cite{GSPK2015} is a slight modification of that in \cite{YMS2003} but the latter involves two lattice nodes.

In this paper, we generalize the idea from \cite{YMS2003,GSPK2015} and construct a family of single-node boundary schemes for the LBM.
The construction are quite universal and simple, it does not involve concrete lattice Boltzmann models and uses the half-way bounce-back rule as a central step.
The boundary schemes thus constructed are all second-order accurate for curved boundaries if so is the bounce-back rule, which is true if the collision term fulfills some simple requirements \cite{ZY} satisfied by many widely used models.
They have good stability thanks to convex combinations.
Furthermore, the second-order accuracy and stability of the schemes are verified by several numerical examples for the multiple-relaxation-time (MRT) models in both 2D and 3D \cite{dHumieres1992rgd,LL2000,DGKLL2002}.
In addition, the constructed schemes contain those in \cite{GSPK2015,ZY2017} as special cases but significantly differ from them.
%

The paper is organized as follows. In Section~\ref{sec2}, we construct a family of single-node
second-order boundary schemes for the LBM.
Some numerical experiments are reported in Section~\ref{sec3} to validate the second-order accuracy and stability of the boundary schemes
for both 2D and 3D MRT models.
Some conclusions and remarks are given in Section~\ref{sec4}.
The paper ends with an appendix for the details of the MRT models used in our numerical experiments.

\section{A family of single-node second-order schemes}
\label{sec2}

The lattice Boltzmann equation (LBE) with general collision models reads as
%
\begin{equation}\label{21}
  f_i (\bm{x} + \bm{e}_i h, \, t+ \delta_t )
  -
  f_i ( \bm{x}, \, t )
  =
  \Omega_i(\bm{x},\, t )   ,   \quad  i=0,1,2,\ldots, q-1.
\end{equation}
Here $f_i(\bm x,t)$ is the $i$-th distribution function for particles with velocity $\bm e_i$ at position $\bm x$ and time $t$;
$h$ and $\delta_t$ are the lattice size and time step, respectively; and $\Omega_i(\bm{x},\, t )$ is the $i$-th collision term.
In the LBE, the discrete velocity set usually satisfies the symmetry $\{ \bm{e}_i \} = \{ - \bm{e}_i \}$ and $\bm{e}_0 = \bm{0}$.
Obviously, the LBE \eqref{21} can be decomposed into the following two steps:
\begin{align}
& f_i^{\prime}(\bm{x},\, t)
  = f_i(\bm{x},\, t) + \Omega_i(\bm{x}, \,  t )  &\quad ( \mbox{collision} ),   \label{22a}\\
& f_i(\bm{x} + \bm e_i h, \, t+ \delta_t)
  =   f_i^{\prime}(\bm{x},\, t)                   &\quad ( \mbox{advection} ).   \label{22b}
\end{align}
It is clear that the collision step is point-wise while the advection step involves two different lattice nodes for $i \neq 0$.

\begin{figure}[!ht]
\begin{center}
\includegraphics[scale=0.5]
{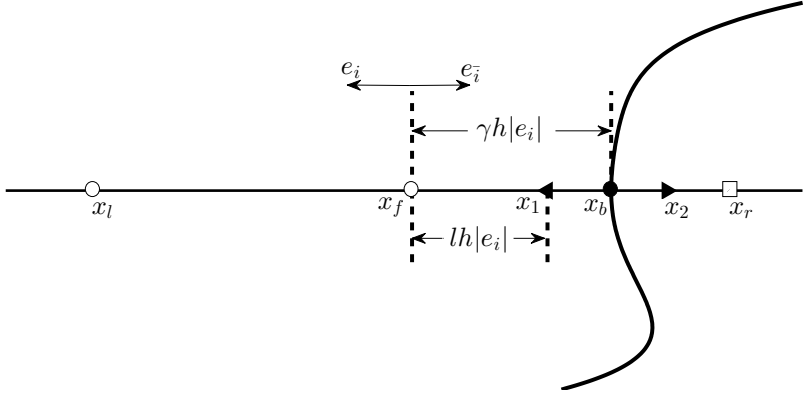}
\caption{
The thin solid straight line is the grid line and the thick curved line is the boundary.
White circles (\Large$\circ$\small) are the fluid nodes, the black circle (\Large$\bullet$\small) is the intersection of the boundary and the grid line,
and the square box ($\square$) is out of the computational domain.
 }
\label{Fig:Interpolation}
\end{center}
\end{figure}

With the above general LBE, we aim at constructing a family of single-node second-order boundary schemes for Dirichlet BCs (see Fig.~\ref{Fig:Interpolation})
\begin{equation} \label{23}
           \bm u ( \bm x , t )
           = \bm{\phi}  ( \bm x , t )
\end{equation}
on the boundary for the incompressible Navier-Stokes equations by generalizing the idea from \cite{YMS2003,GSPK2015}.
Here $\bm u ( \bm x , t )$ is the macroscopic fluid velocity at position $\bm{x}$ and time $t$,
 $\bm{\phi}  ( \bm x , t )$ is a given function of $\bm{x}$ and $t$, and the boundary
is often curved in complex flows (\eg, flows in porous media \cite{Chai2016} and multi-phase flows \cite{Fakhari2017}).

For the sake of definiteness, we fix the direction $\bm e_i$ and construct a formula to compute the distribution $f_i(\bm x_f, t+\delta_t)$ at the lattice node $\bm x_f$ next to the boundary as illustrated in Fig.~\ref{Fig:Interpolation}.
Denote by $\bm x_b, \bm x_l$ and $\bm x_r$ the intersection of the given boundary and the grid line in the $\bm e_i$-direction, and the left and right neighboring lattice nodes of $\bm x_f$.
Namely,
$$
\bm x_l = \bm x_f + h{\bm e}_{i}, \qquad
\bm x_r = \bm x_f - h {\bm e}_{i},
$$
$$
\bm x_b = \bm x_f - \gamma h {\bm e}_{i}, \qquad \gamma\in(0, 1].
$$
Additionally, let $l$ be a non-negative number and take
$$
\bm x_1 = \bm x_f - lh{\bm e}_{i}, \qquad
\bm x_2 = 2\bm x_b - \bm x_1
$$
(See Fig.~\ref{Fig:Interpolation}).

With $\bm x_1$ and $\bm x_2$ defined above, we firstly interpolate the distribution function $f_i(\bm x_f, t+\delta_t)$ with those at $\bm x_l$ and $\bm x_{1}$ by
\begin{equation*}
f_i(\bm x_f, t+\delta_t) = \frac{l}{1+l} f_i(\bm x_l, t+\delta_t) +  \frac{1}{1+l} f_i(\bm x_{1}, t+\delta_t).
\end{equation*}
Notice that $l \geq 0$. Thanks to the advection $f_i(\bm x_l, t+\delta_t) = f^{\prime}_i(\bm x_f, t)$, the above can be rewritten as
\begin{equation}\label{25}
f_i(\bm x_f, t+\delta_t) = \frac{l}{1+l} f^{\prime}_i(\bm x_f, t) +  \frac{1}{1+l} f_i(\bm x_{1}, t+\delta_t).
\end{equation}

For $f_i(\bm x_{1}, t+\delta_t)$ in \eqref{25}, we compute it with the well-known half-way bounce-back scheme \cite{Ladd1, Ladd2} (the boundary point $\bm x_b$ is located at the middle of $\bm x_1$ and $\bm x_2$!)
\begin{equation}\label{26}
f_i(\bm x_1, t+\delta_t) = f_{\bar i}(\bm x_2, t+\delta_t) + 2 \omega_i h\rho_{0}\frac{\bm{e}_{i}\cdot\bm{\phi}( \bm x_b, t ) }{\delta_tc_{s}^{2}}.
\end{equation}
Here $\bar i$ is such that $\bm e_{\bar i} = - \bm e_i$ and the constants $\omega_i, \rho_0$ and $c_s$ are explained in Appendix (see also \cite{Ladd1, Ladd2}).

It remains to compute $f_{\bar i}(\bm x_2, t+\delta_t)$ in \eqref{26}. We interpolate it with the distribution functions at $\bm x_{f}$ and $\bm x_r$:
\begin{equation*}
f_{\bar i} (\bm x_2, t+\delta_t) = (1+l-2\gamma) f_{\bar i} (\bm x_f, t+\delta_t)
                                   +  (2\gamma-l) f_{\bar i} (\bm x_r, t+\delta_t).
\end{equation*}
Again, we use the advection $f_{\bar i} (\bm x_r, t+\delta_t)= f^{\prime}_{\bar i} (\bm x_f, t)$ to obtain
\begin{equation}\label{27}
f_{\bar i} (\bm x_2, t+\delta_t) = (1+l-2\gamma) f_{\bar i} (\bm x_f, t+\delta_t)
                                   +  (2\gamma-l) f^{\prime}_{\bar i} (\bm x_f, t).
\end{equation}
Combining Eqs.~\eqref{25}--\eqref{27} gives
\begin{equation}\label{28}
\begin{split}
f_{i}(\bm x_{f},t+\delta_{t})
 = &   \frac{1+l-2\gamma}{1+l}    f_{\bar i}(\bm x_{f},t+\delta_t)
       + \frac{l}{1+l}            f_{i}^{\prime}(\bm x_{f},t) \\
   &   + \frac{2\gamma- l}{1+l}   f_{\bar i}^{\prime}(\bm x_{f},t)
       + \frac{2}{1+l}             \omega_{i}h\rho_{0}\frac{\bm{e}_{i}\cdot\bm{\phi}( \bm x_b, t) }{\delta_t c_{s}^{2}}.
\end{split}
\end{equation}
Furthermore, with the approximation
\begin{equation}\label{29}
f_{\bar i}(\bm x_{f},t+\delta_t) \approx f_{\bar i}(\bm x_{f},t)
\end{equation}
in Eq.~\eqref{28}, we arrive at the following single-node scheme
\begin{equation}\label{210}
\begin{split}
f_{i}(\bm x_{f},t+\delta_{t})
 =  &  \frac{1+l-2\gamma}{1+l}    f_{\bar i}(\bm x_{f},t)
       + \frac{l}{1+l}            f_{i}^{\prime}(\bm x_{f},t) \\
    &  + \frac{2\gamma- l}{1+l}   f_{\bar i}^{\prime}(\bm x_{f},t)
       + \frac{2}{1+l}             \omega_{i}h\rho_{0}\frac{\bm{e}_{i}\cdot\bm{\phi} ( \bm x_b, t) }{\delta_t c_{s}^{2}}
\end{split}
\end{equation}
parameterized with $l\geq 0$.

About this scheme, we have the following remark.

\textbf{Remark. }
{\it
(1). The above construction is quite universal, it relies only on the half-way bounce back rule but does not involves the specific form of the collision term.

(2). In case that the left lattice node $\bm x_l$ belongs to the computational domain which is often true, we can replace the approximation Eq.~\eqref{29} with
$f_{\bar i}(\bm x_{f},t+\delta_t)=f_{\bar i}^{\prime}(\bm x_l,t)$ in Eq.~\eqref{28} to obtain the following two-node scheme
\begin{equation}\label{211}
\begin{split}
f_{i}(\bm x_{f},t+\delta_{t})
 =  &  \frac{1+l-2\gamma}{1+l}    f_{\bar i}^{\prime}(\bm x_l,t)
       + \frac{l}{1+l}            f_{i}^{\prime}(\bm x_{f},t) \\
    &  + \frac{2\gamma- l}{1+l}   f_{\bar i}^{\prime}(\bm x_{f},t)
       + \frac{2}{1+l}             \omega_{i}h\rho_{0}\frac{\bm{e}_{i}\cdot\bm{\phi}( \bm x_b, t ) }{\delta_t c_{s}^{2}}.
\end{split}
\end{equation}
(3). In both \cite{YMS2003} and \cite{GSPK2015}, the point $\bm x_1$ is chosen as $\bm x_1 = \bm x_b$, namely, $l=\gamma$ and $\bm x_2 = \bm x_1$. Here we choose $\bm x_1$ quite arbitrarily and thus obtain a family of boundary schemes.

(4). When $l=\gamma$, Scheme \eqref{211} degenerates to the non-single-node scheme proposed in \cite{YMS2003}:
\begin{equation*}\label{311-1}
f_{i}(\bm x_{f},t+\delta_{t})
    =  \frac{1-\gamma}{1+\gamma} f_{\bar i}^{\prime}(\bm x_l,t)
       + \frac{\gamma}{1+\gamma}
         \left[ f_{i}^{\prime}(\bm x_{f},t)
                + f_{\bar i}^{\prime}(\bm x_{f},t)
         \right]
       + \frac{2}{1+\gamma}\omega_{i}h\rho_{0}\frac{\bm{e}_{i}\cdot\bm{\phi} ( \bm x_b, t ) }{\delta_t c_{s}^{2}},
\end{equation*}
while \eqref{210} becomes that in \cite{GSPK2015}:
\begin{equation*}\label{311-2}
f_{i}(\bm x_{f},t+\delta_{t})
    =  \frac{1-\gamma}{1+\gamma} f_{\bar i}(\bm x_{f},t)
       + \frac{\gamma}{1+\gamma}
         \left[ f_{i}^{\prime}(\bm x_{f},t)
                + f_{\bar i}^{\prime}(\bm x_{f},t)
         \right]
       + \frac{2}{1+\gamma}\omega_{i}h\rho_{0}\frac{\bm{e}_{i}\cdot\bm{\phi} ( \bm x_b, t ) }{\delta_t c_{s}^{2}}.
\end{equation*}

(5). When $l=0$ and $2 \gamma$, Scheme \eqref{210} degenerates to our nonconvex and convex schemes proposed in \cite{ZY2017}, respectively.
}\\

The second-order accuracy of the single-node scheme \eqref{210} can be simply explained as follows.
First, two interpolations \eqref{25} and \eqref{27} are second-order accurate.
In addition, for the diffusive scaling $\delta_t = \eta h^2$ ($\eta$ is an adjustable parameter),
the approximation \eqref{29} is of $O(h^2)$.
Moreover, assume that the half-way bounce-back rule \eqref{26} has second-order accuracy, which is true if the collision term fulfills some simple requirements \cite{ZY} satisfied by many widely used models. Therefore the scheme \eqref{210} is second-order accurate.

Next we discuss the stability of the scheme \eqref{210}.
To ensure the stability of interpolations \eqref{25} and \eqref{27}, we require that the interpolation coefficients belong to $[0,1]$, $i.e.$,
$$
l \geq 0, \quad 1-2\gamma+l \geq 0 \quad \mbox{and} \quad 2\gamma-l \geq 0 .
$$
Namely,
\begin{equation}\label{212}
\max\{0, 2\gamma -1\}\leq l\leq 2\gamma .
\end{equation}
These are exactly the conditions ensuring that the scheme  \eqref{210} is a convex combination of the distribution functions.

Finally, we notice that Scheme \eqref{210} does not involve the distribution $f_{i}(\bm x_{f},t)$. Thus, we may propose a more general boundary scheme by replacing the right-hand side of Scheme \eqref{210} with a convex combination of $f_{i}(\bm x_{f},t)$ and the right-hand side:
\begin{equation}\label{213}
\begin{split}
f_{i}(\bm x_{f},t+\delta_{t})
 & = (1-b) f_{i}(\bm x_{f},t)
       + b\left[ \frac{1+l-2\gamma}{1+l}    f_{\bar i}(\bm x_{f},t)+ \frac{l}{1+l}            f_{i}^{\prime}(\bm x_{f},t) \right.\\[3mm]
       &\left.\qquad \qquad
      + \frac{2\gamma- l}{1+l}   f_{\bar i}^{\prime}(\bm x_{f},t)
       + \frac{2}{1+l}             \omega_{i}h\rho_{0}\frac{\bm{e}_{i}\cdot\bm{\phi} ( \bm x_b, t) }{\delta_t c_{s}^{2}}\right].
\end{split}
\end{equation}
This new scheme contains two free parameters $l$ and $b\in(0,1]$. Since we use the diffusive scaling, the approximation of $f_{i}(\bm x_{f},t+\delta_{t})$ by $f_{i}(\bm x_{f},t)$ is second-order accurate.
Therefore, the new scheme \eqref{213} has second-order accuracy too.

\section{Numerical experiments}
\label{sec3}

In this section, we report several numerical experiments to validate the single-node boundary scheme \eqref{210}.
Since this scheme contains an adjustable parameter $l$ satisfying the constraints in \eqref{212}, there are infinitely many boundary schemes.
To be concrete, we will restrict ourselves to the following five cases: $l=\gamma,\  1.5\gamma, \ 2\gamma, \ \gamma^2$ and $\gamma^2+\gamma$.
%
%
%

On the other hand, we will only consider the widely used D2Q9 and D3Q15 multiple-relaxation-time (MRT) models \cite{dHumieres1992rgd,LL2000,DGKLL2002},
whose details are given in Appendix.
For these two MRT models, there are infinitely many choices of relaxation rates.
In the simulations, we only change the relaxation rate $s_{\nu}$ related to the viscosity and fix all the others to examine the accuracy and stability of the schemes. Without loss of generality, we take the relaxation rates for the D2Q9 model as
\begin{equation}
\mathsf{S}= \diag( 1, 1.8, 1.2, 1, 0.5, 0.5, 1, s_{\nu}, s_{\nu}  )
\end{equation}
and
\begin{equation}
\mathsf{S}= \diag( 1, 1.8, 1.2, 1, 0.5,1, 0.5,1,0.5, s_{\nu}, s_{\nu}, s_{\nu}, s_{\nu}, s_{\nu}, 1.5 )
\end{equation}
for the D3Q15 model.
Recall that we use the diffusive scaling $\delta_t = \eta h^2$.
Then the relations between $s_{\nu}$ and the kinematic viscosity $\nu$  for the above two models are both
\begin{equation}
\nu = \frac{1}{3\eta}( \frac{1}{s_{\nu}} - \frac{1}{2} )
\end{equation}
(see \cite{dHumieres1992rgd,LL2000,DGKLL2002}).
From this, $\eta$ can be determined via $\nu$ and $s_{\nu}$.

With the above choice of parameters, we conduct numerical experiments for the following three problems: the Poiseuille flow with straight boundaries, the Taylor-Green vortex flow with curved boundaries, and the 3D Hagen-Poiseuille flow in a circular pipe.
All these flows are governed by the incompressible Navier-Stokes equations
\begin{align}  \label{eqn:NS}
\nabla\cdot{\bm u}=0,\qquad \partial_t {\bm u} + {\bm u}\cdot\nabla{\bm u} + \nabla p = \nu\Delta{\bm u} + \bm F
\end{align}
in proper domains, where $\nu$ is the kinematic viscosity and $\bm F$ is an external force. They all have analytical solutions. For each numerical experiment, we only need to specify the relaxation rate $s_{\nu}$ and lattice size $h$, which determine all other parameters: $\delta_t = \eta h^2$ and $\eta = (1/{s_{\nu}} - 1/2)/(3\nu)$.

\subsection{Poiseuille flow}

\begin{figure}[!ht]
\begin{center}
\includegraphics[scale=0.4]{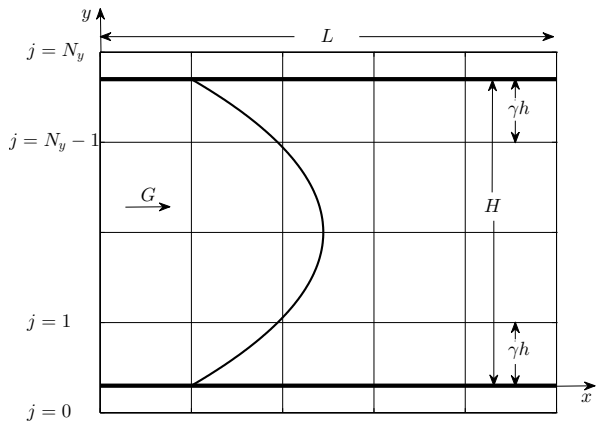}
\caption{Configuration of the Poiseuille flow in LBE simulations with an arbitrary $\gamma$. }
\label{Fig:Poiseuille}
\end{center}
\end{figure}

The first problem is the Poiseuille flow between two parallel no-slip
walls driven by a constant body force $\bm F = G(1, 0)$ (see Fig.~\ref{Fig:Poiseuille}). This problem has the following analytical solution
\begin{equation}\label{52}
        u = u(y)=4 U  ( 1-\frac{y}{H} ) \frac{y}{H} ,  \quad  v = 0,
\end{equation}
for $y\in[0, H]$.
Here $(u, v) = {\bm u}$, $H$ is the channel width, $U = G H^2 /8\nu$ is the maximal velocity along the center line of the channel, and the parameters are
\begin{equation*}
          \nu=0.03, \quad  G = 0.8\nu, \quad H=1.
\end{equation*}

In our computation, the horizontal direction is periodic.
The boundary schemes are applied at the upper and lower straight boundaries.
As illustrated in Fig.~\ref{Fig:Poiseuille}, $N_y$ is the number of meshes in the vertical direction, and the lower and upper walls are located between $j=0$ and $j=1$, $j=N_y$ and $j=N_y-1$, respectively.
The lattice size is
\begin{equation}\label{53}
   h = \frac{H}{N_y-2+2\gamma}
\end{equation}
with $\gamma$ the scaled distance.
To demonstrate the accuracy and stability of the boundary schemes, we define the relative $L^2$-error as
\begin{equation} \label{eqn:L2-error}
        E_r=
        \frac {\sqrt{ \sum _{\bm x } \vert \bm u(\bm x ) - \bm u^{*} (\bm x ) \vert ^2 }}
        {\sqrt{ \sum _{\bm x } \vert  \bm u(\bm x )  \vert ^2}}
        ,
\end{equation}
where the summation is over all lattice nodes in the computational domain, $\bm u=(u, v)$ is the analytical solution \eqref{52}, and $\bm u^{*}$ is the LB solution.

\begin{figure}[!ht]
\begin{center}
\includegraphics[scale=0.3]{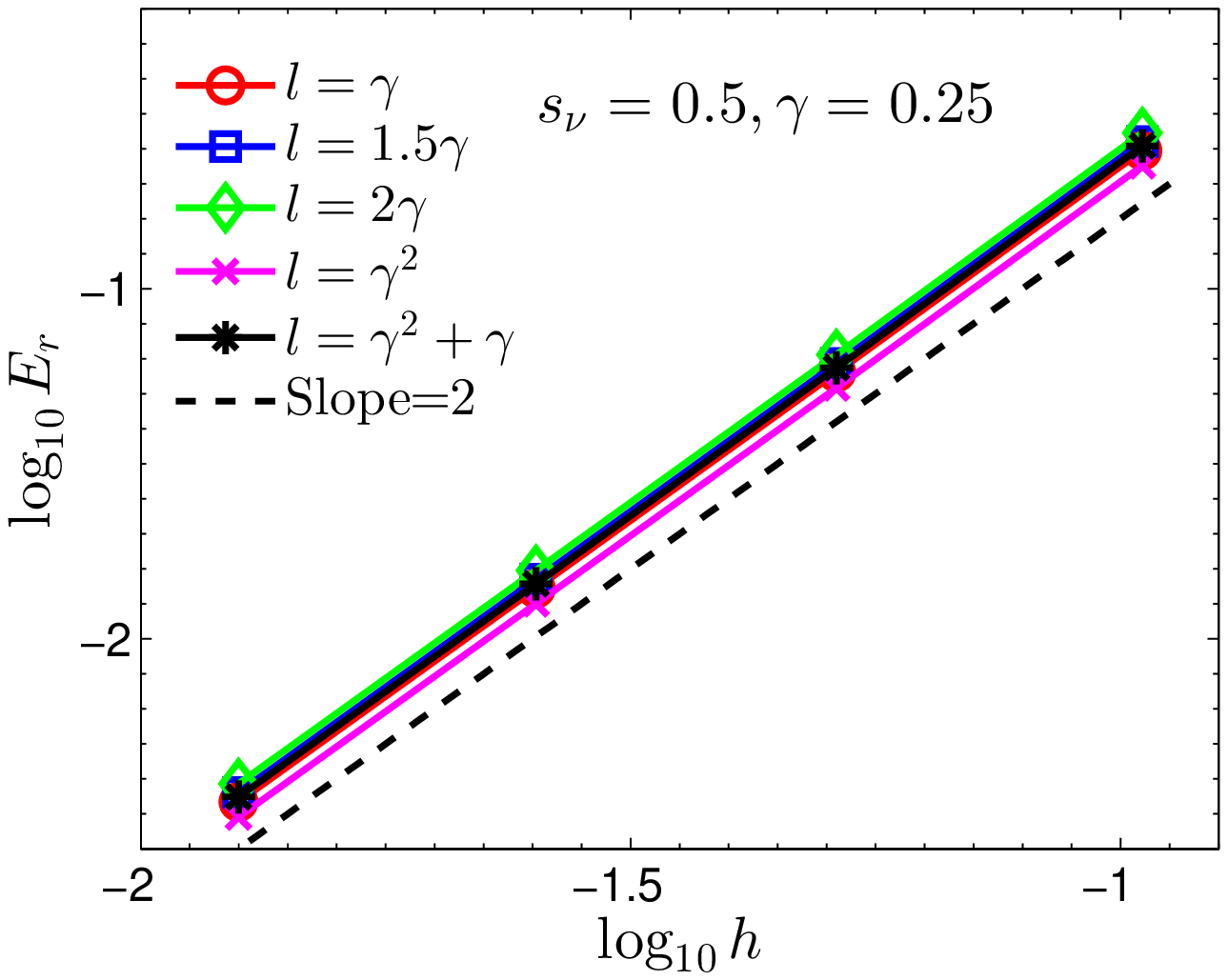}
\includegraphics[scale=0.3]{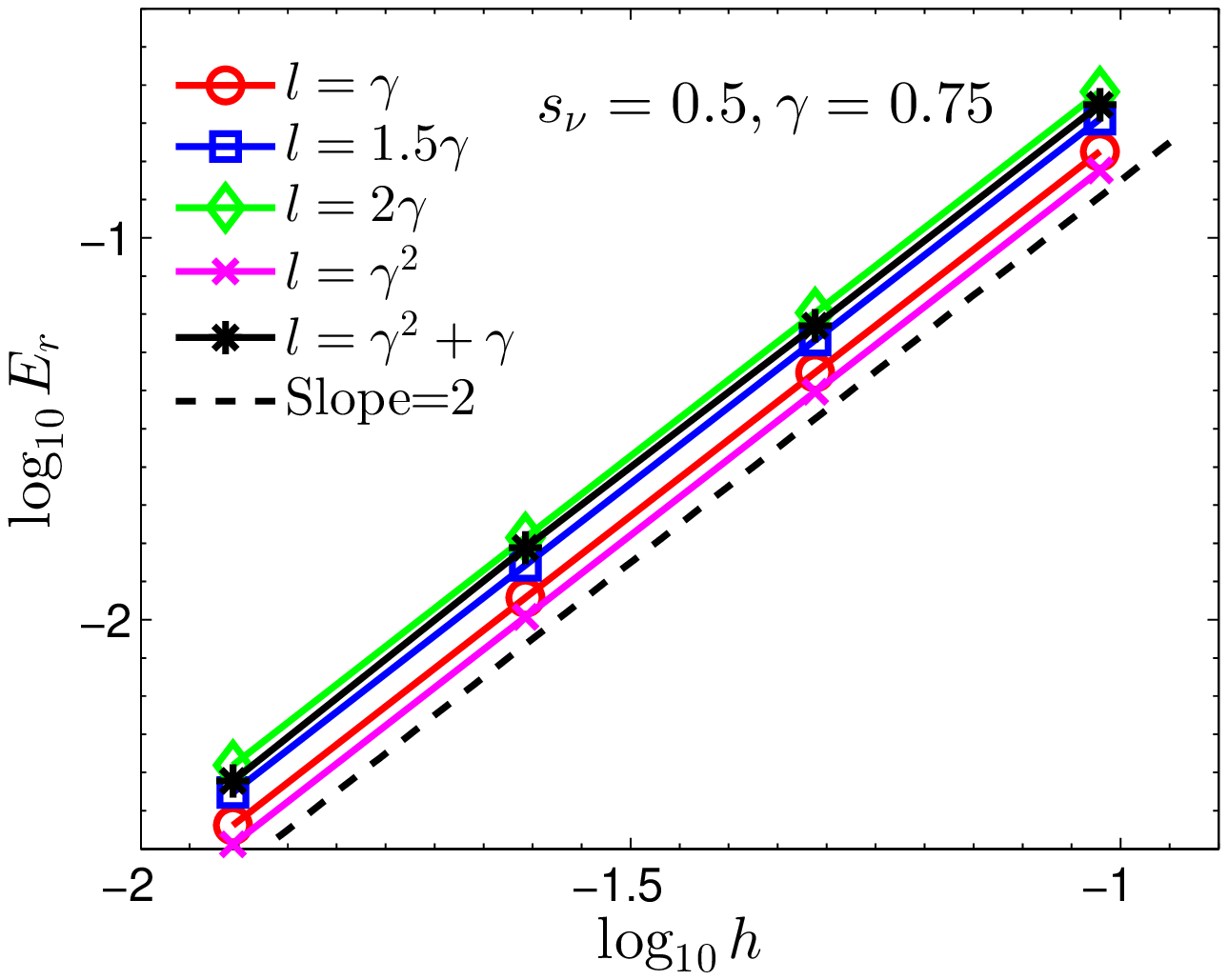}
\includegraphics[scale=0.3]{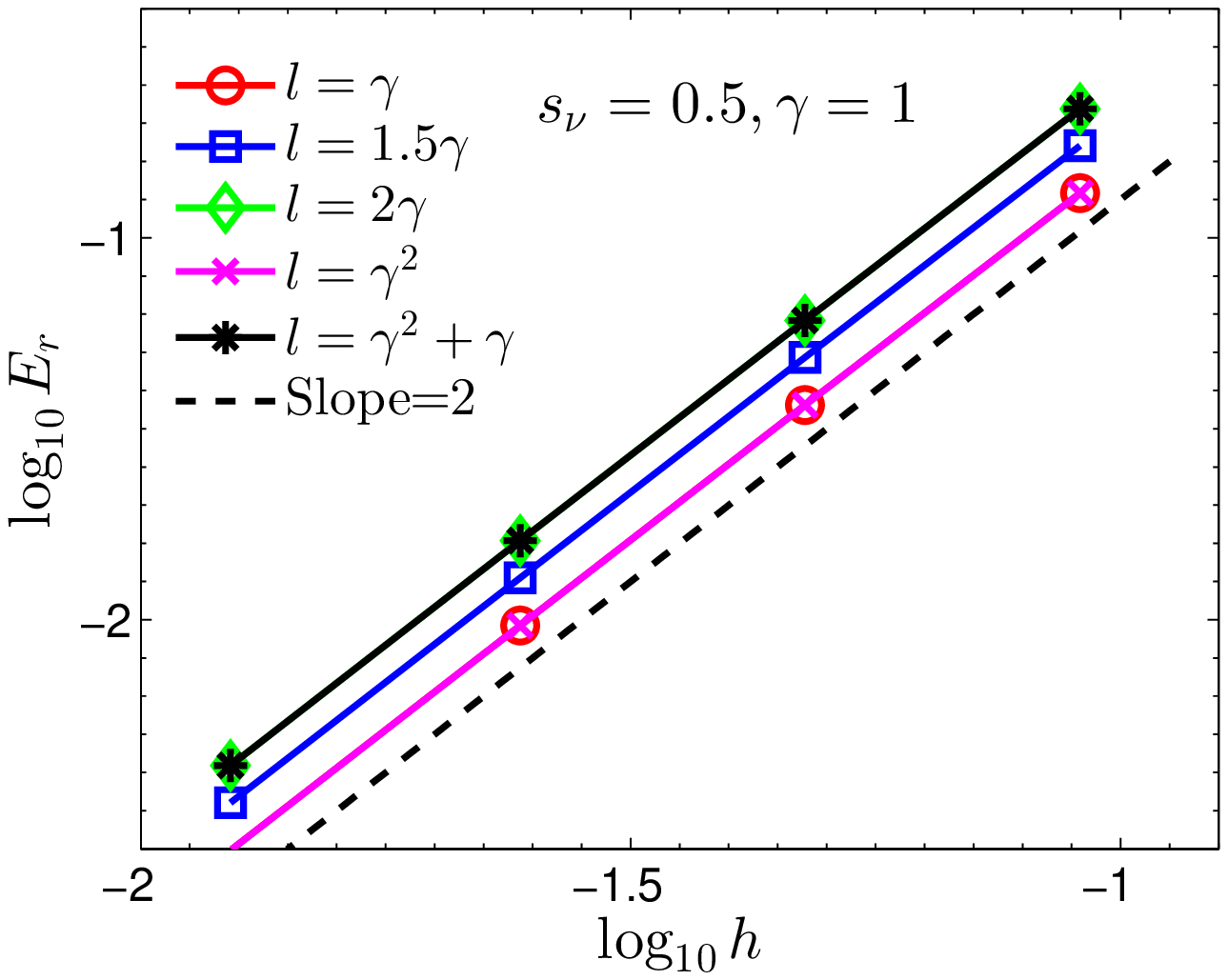}
\includegraphics[scale=0.3]{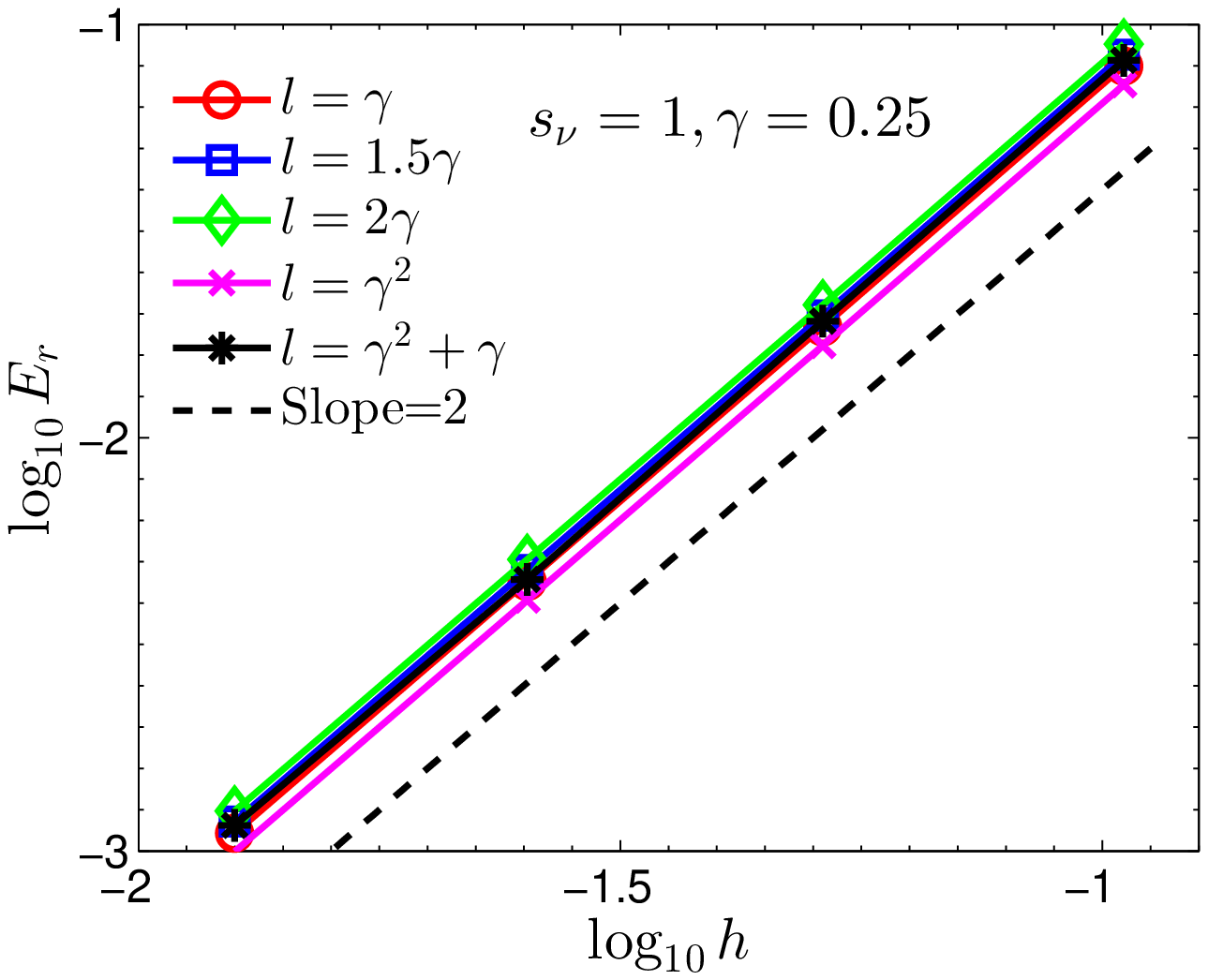}
\includegraphics[scale=0.3]{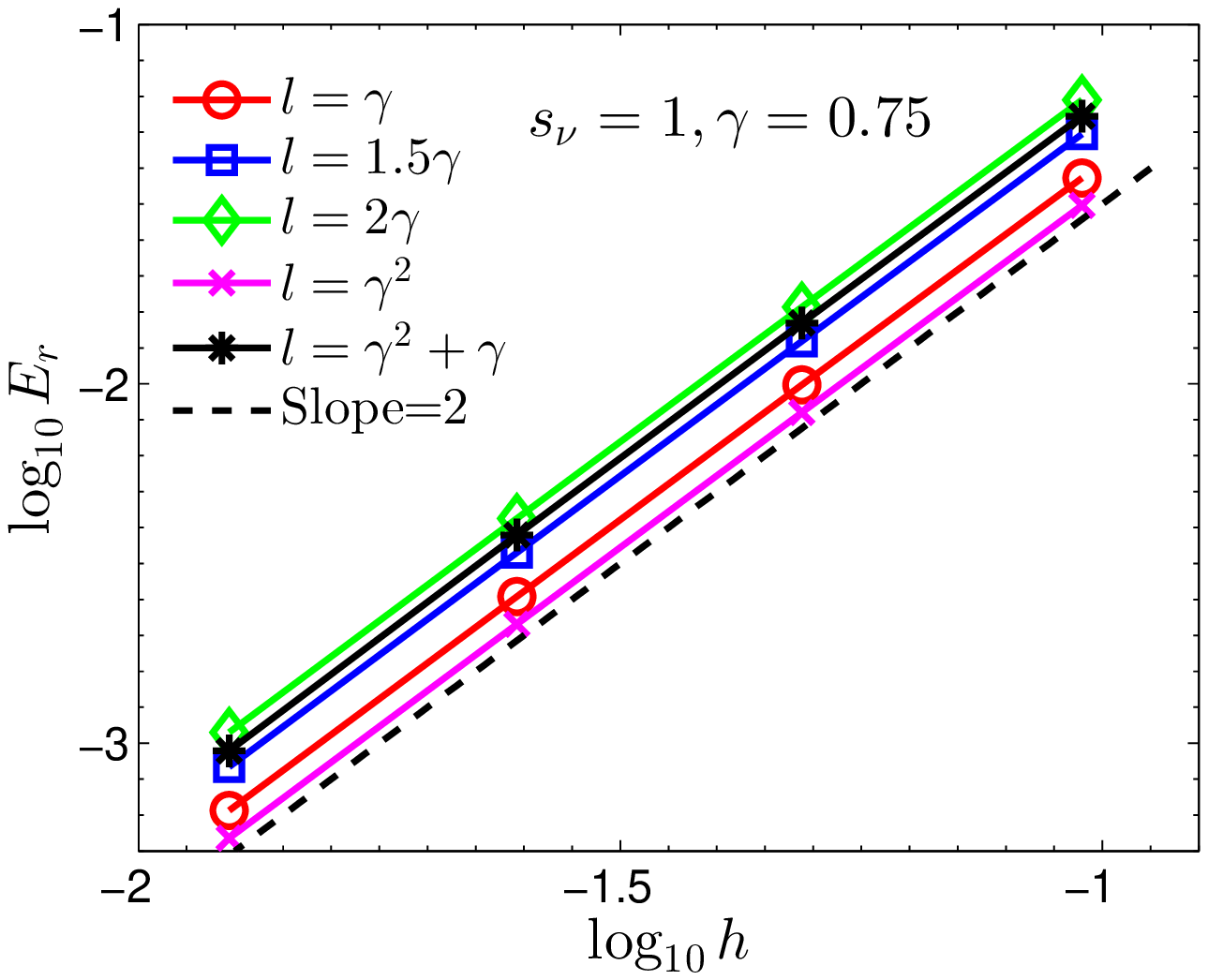}
\includegraphics[scale=0.3]{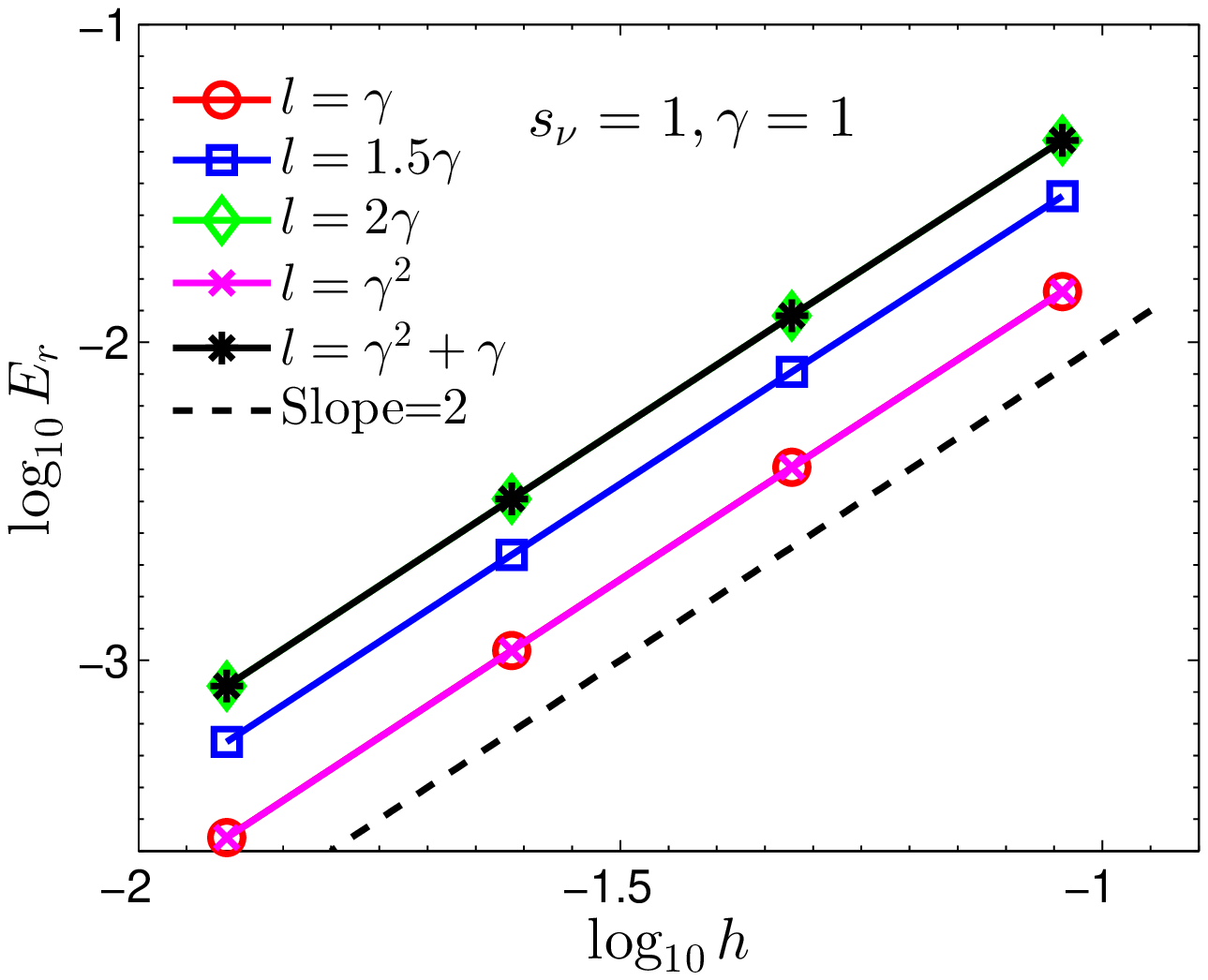}
\includegraphics[scale=0.3]{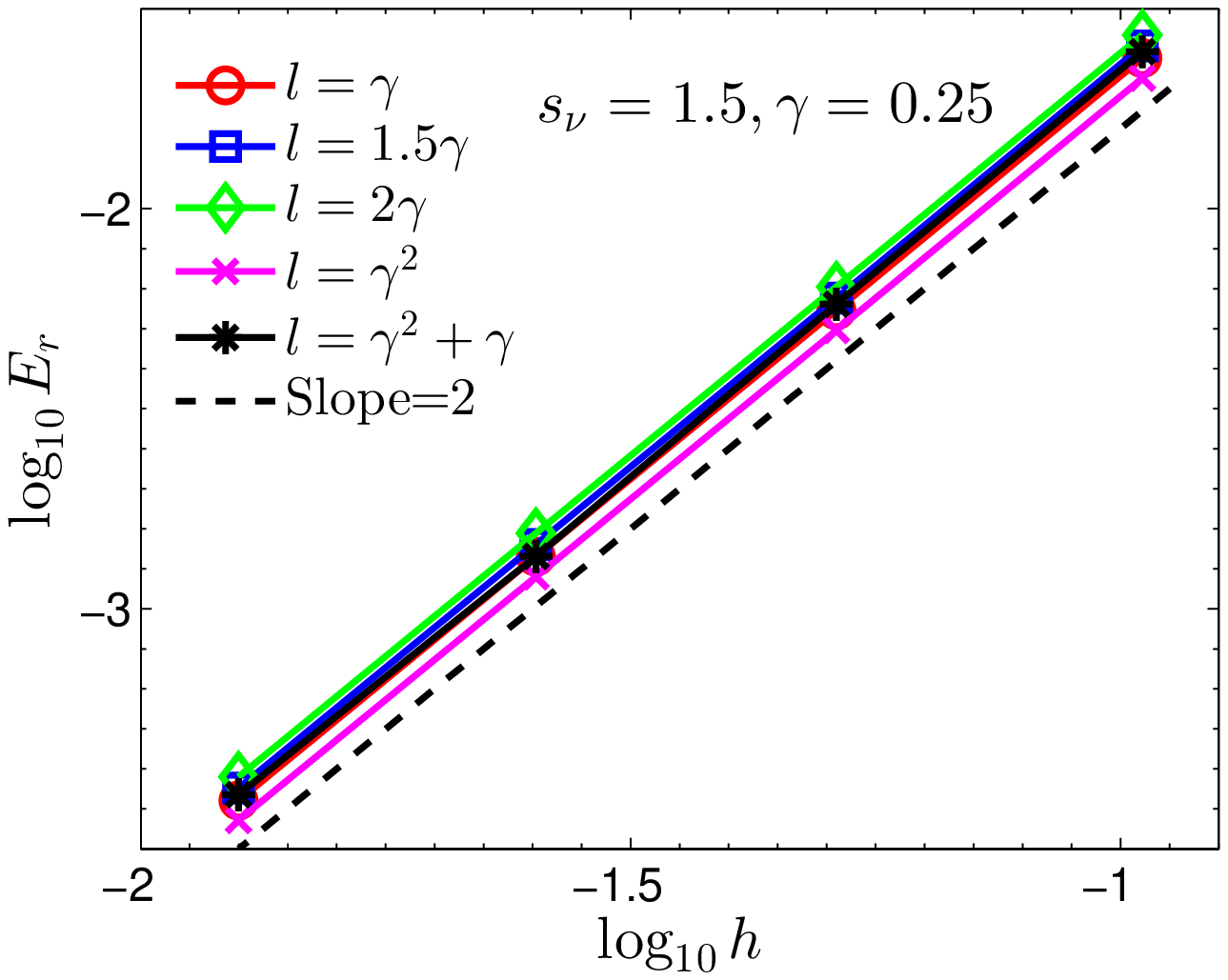}
\includegraphics[scale=0.3]{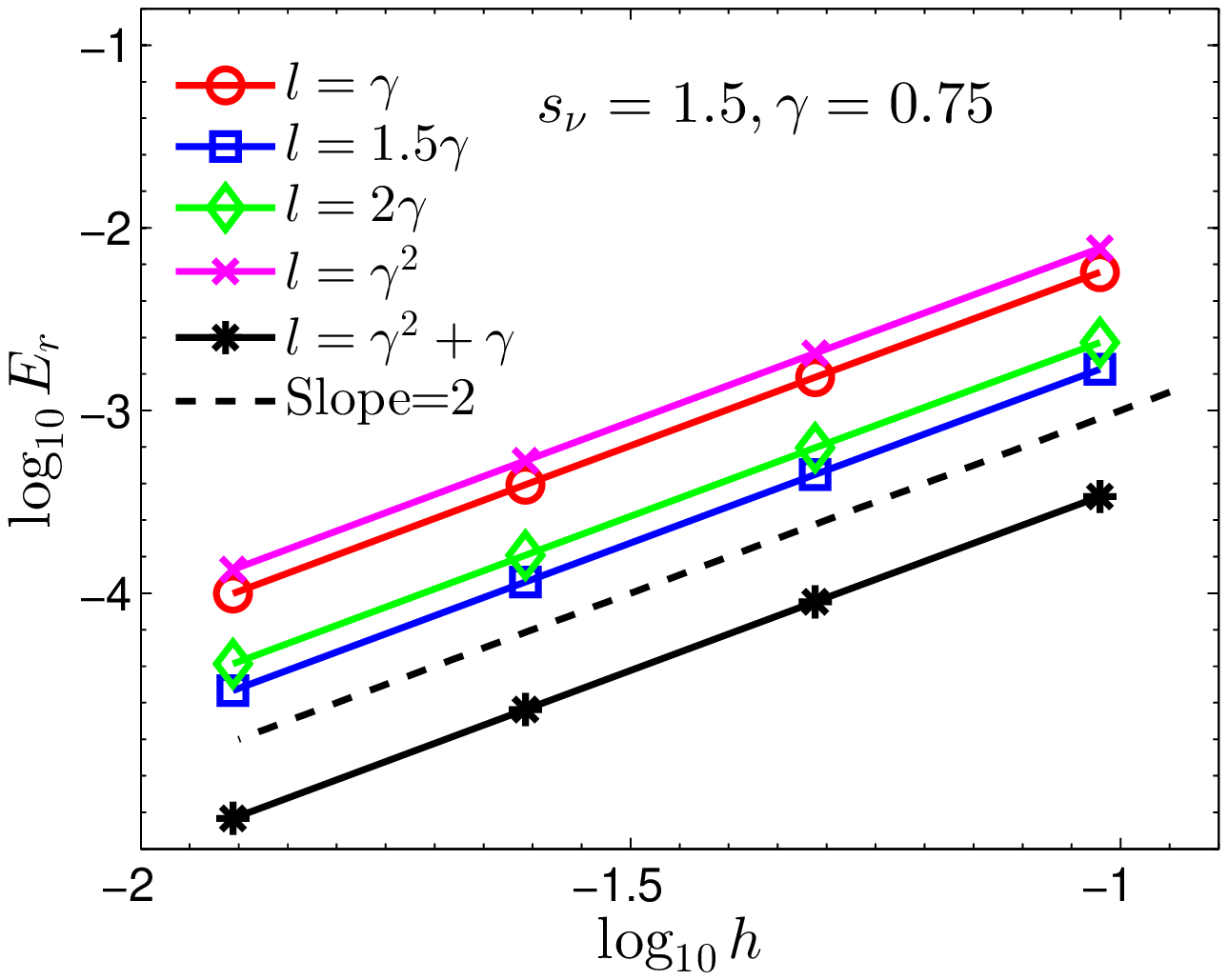}
\includegraphics[scale=0.3]{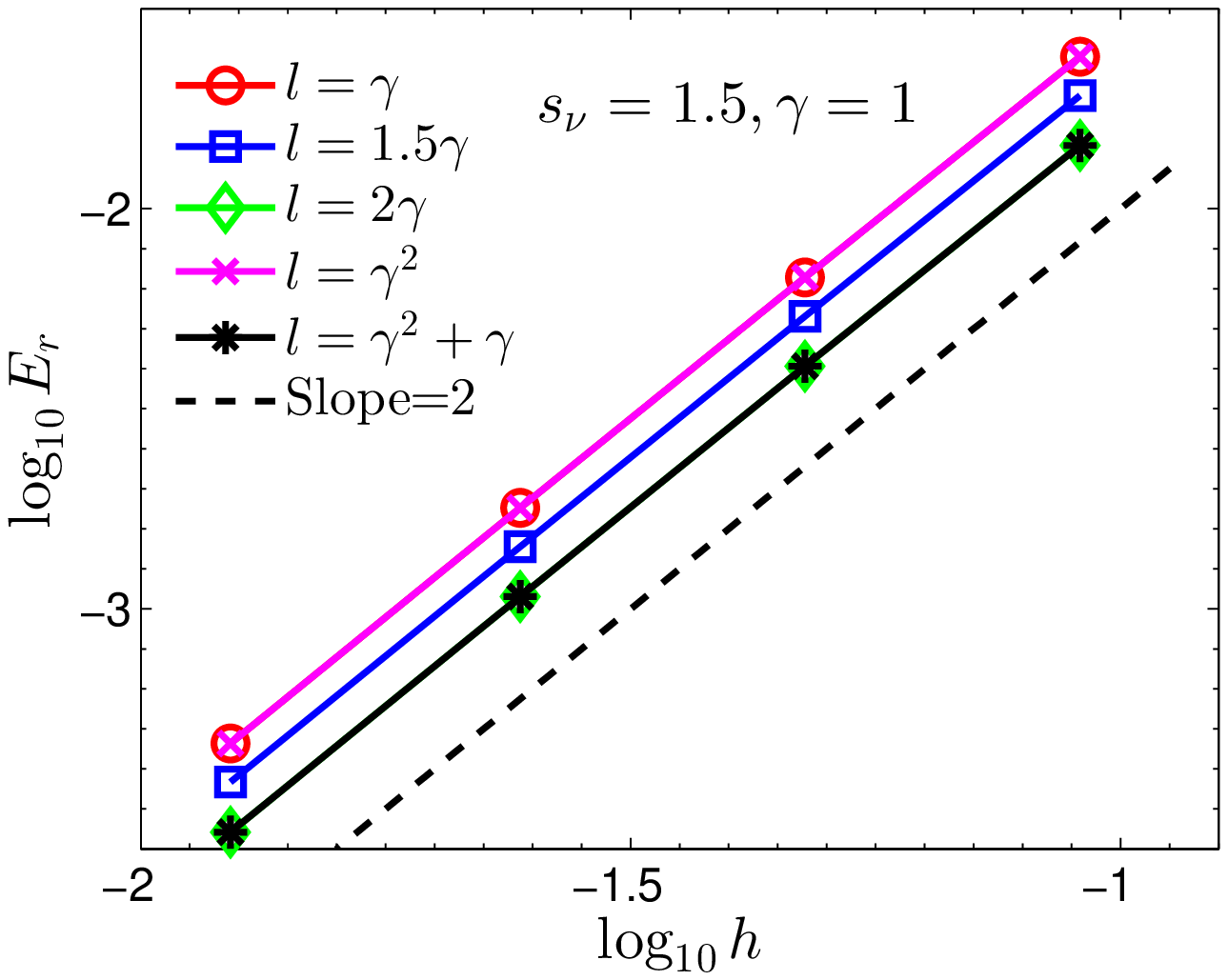}
\includegraphics[scale=0.3]{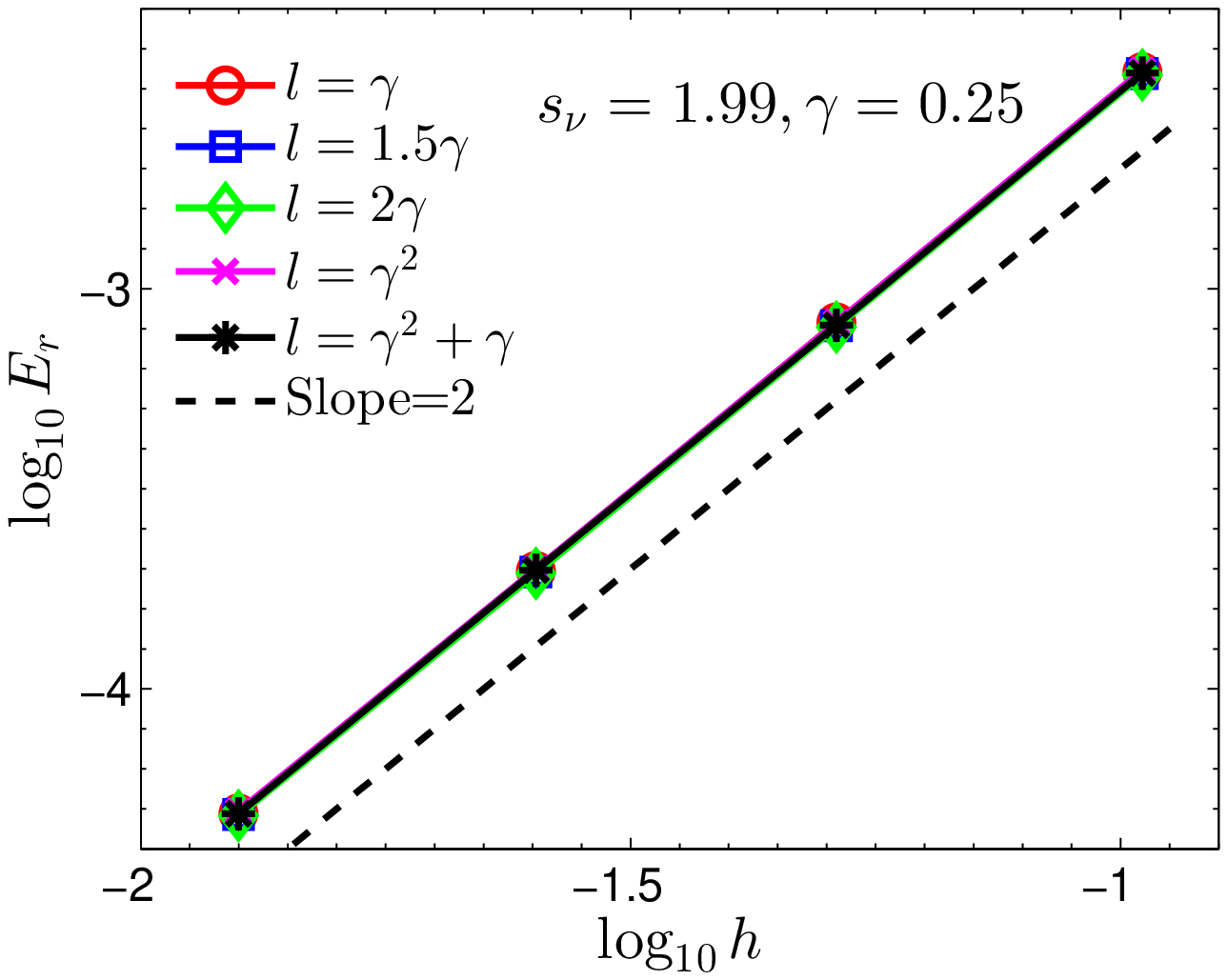}
\includegraphics[scale=0.3]{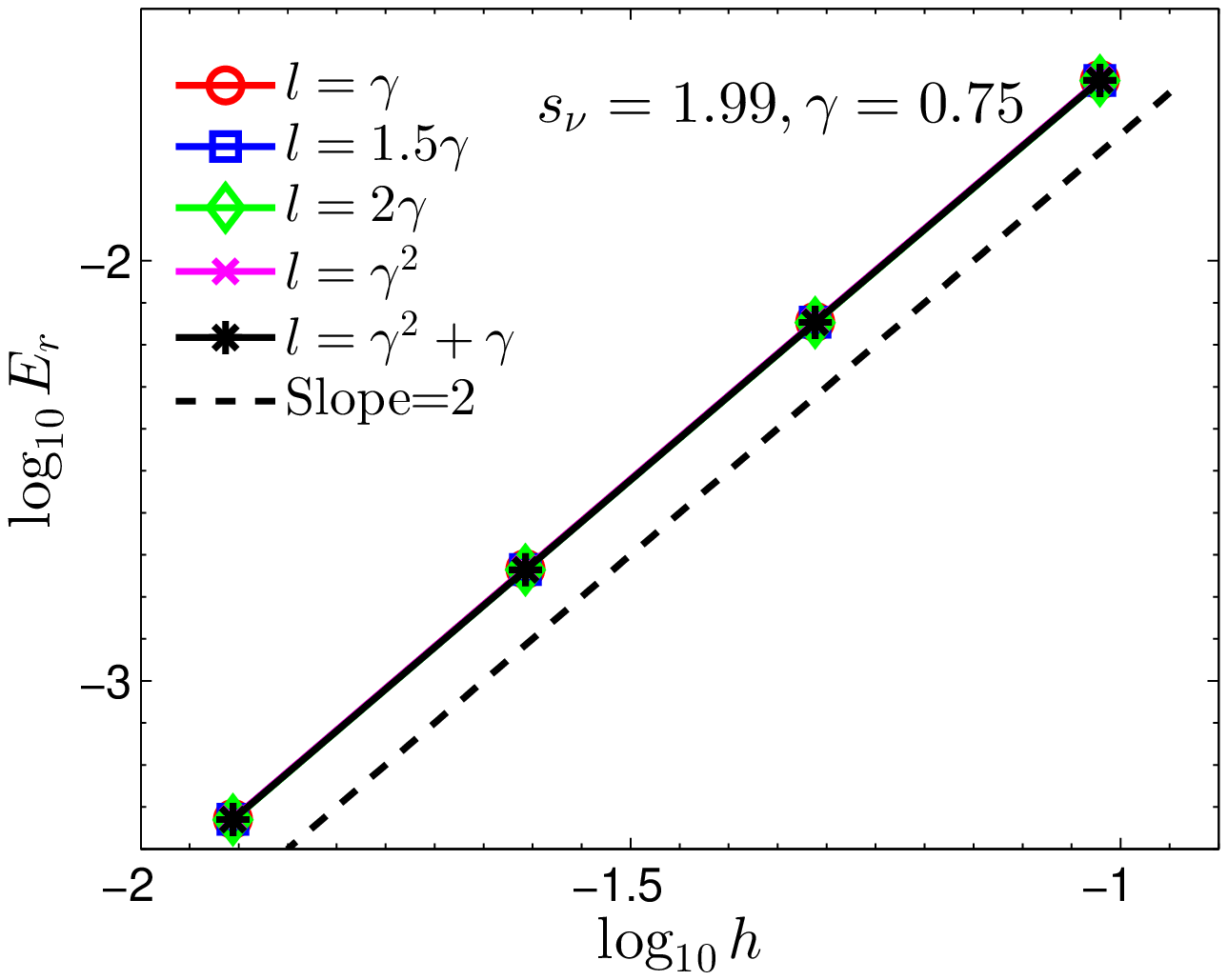}
\includegraphics[scale=0.3]{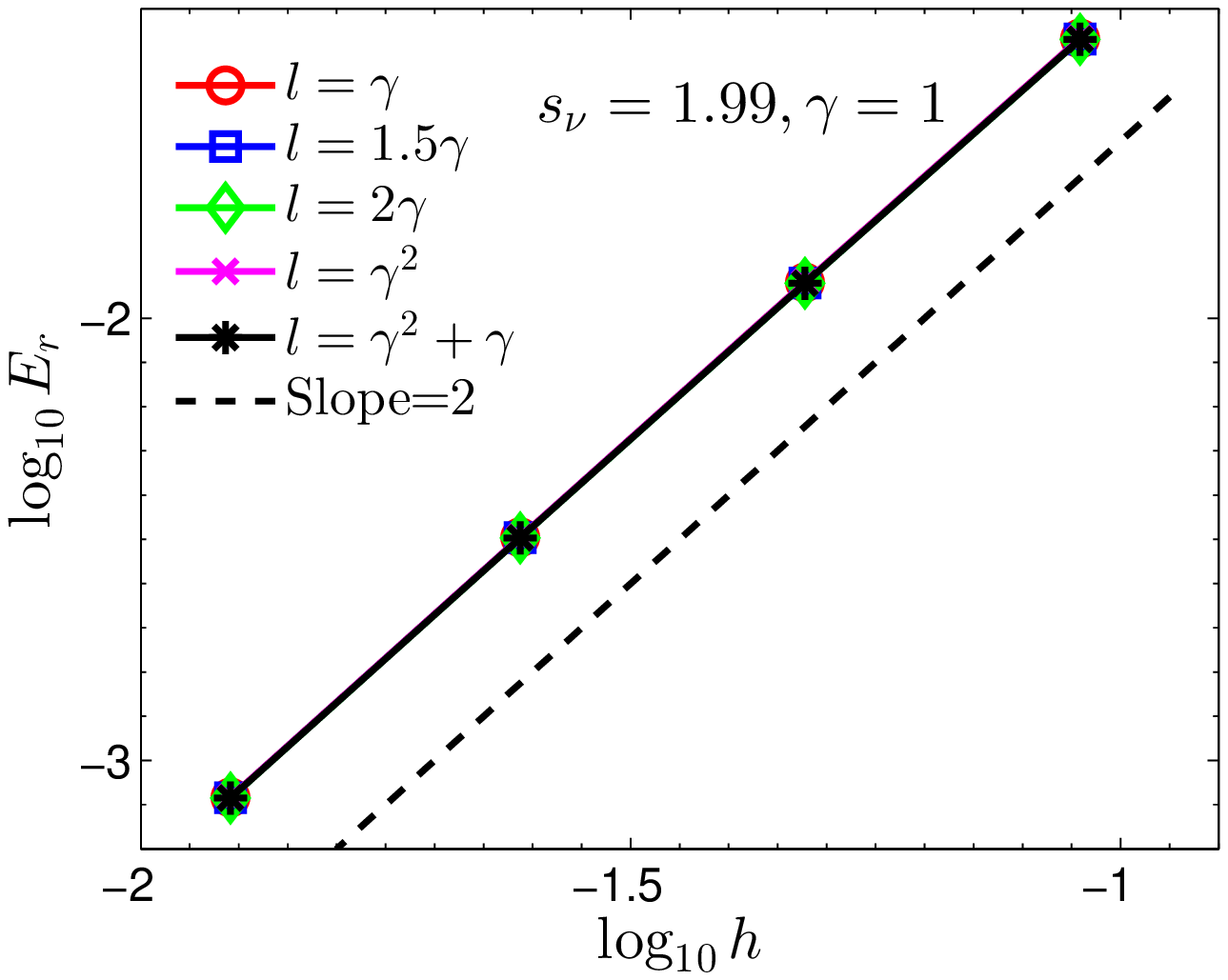}
\caption{Convergence order of the boundary schemes for the Poiseuille flow.
From left to right: $\gamma=0.25$, $0.75$ and $1$. From top to bottom: $s_{\nu}=0.5,1,1.5,1.99$. }
\label{Fig:ConvergenOfPoiseuille}
\end{center}
\end{figure}

In our numerical experiments, we set $\gamma=0.25, 0.75$ and $1$, take different $s_{\nu}$ (=0.5,1,1.5,1.99) and $N_y=11,21,41,61,81$, and the number of meshes in the horizontal direction is $N_x=2(N_y-1)$.
Note that the lattice size $h$ is calculated by Eq.~\eqref{53}.
Fig.~\ref{Fig:ConvergenOfPoiseuille} shows that the convergence orders are around 2 for all the five schemes with different $\gamma$ and $s_{\nu}$ .
These show the second-order accuracy of the five schemes for straight boundaries.

\subsection{Taylor-Green vortex flow in a circular domain}

The second problem we consider is the Taylor-Green vortex flow
in the circular domain
\begin{align*}  \label{eqn:Taylor-Green-solution}
      \Omega:=\left\{ (x,y) \vert (x-\frac{1}{2})^2+(y-\frac{1}{2})^2 \leq \frac{1}{16} \right \}
\end{align*}
without external forces. This problem has analytic solutions
\begin{equation*}
\begin{split}
      & u = -U_0  \cos (2\pi x) \sin (2\pi y) e^{-8 \pi^2 \nu t}, \\
      & v = U_0 \cos (2 \pi y) \sin (2\pi x) e^{-8 \pi^2 \nu t},\\
      & p = p_0 - \frac{1}{4} U_0^2 \left[ \cos (4\pi x) + \cos (4\pi y) \right]e^{-16 \pi^2 \nu t}
\end{split}
\end{equation*}
with free parameters $U_0$ and $p_0$. In our numerical simulations reported below, we take the parameters as
\begin{equation*}
         \nu=0.002, \quad U_0 = 0.05,  \quad  p_0= \rho_0 c_s^2 \quad \mbox{with} \quad \rho_0 =1.
\end{equation*}
The initial and boundary values are given by the above analytical solutions.

Let $\bm u^*=\bm u(\bm x, t)$ be the LB solution and $\bm u=(u, v)$ the above analytic solution.
We define the relative $L^2$-error as
\begin{equation} \label{eqn:L2-error-2}
        E_r=
        \frac {\sqrt{ \sum _{\bm x} \vert \bm u(\bm x, T ) - \bm u^{*} (\bm x, T  ) \vert ^2 }}
        {\sqrt{ \sum _{\bm x} \vert  \bm u(\bm x, T )  \vert ^2}}
\end{equation}
at time $T=1/U_0$,
where the summation is over all lattice nodes in the circular domain $\Omega$.

\begin{figure}[!ht]
\begin{center}
\includegraphics[scale=0.4]{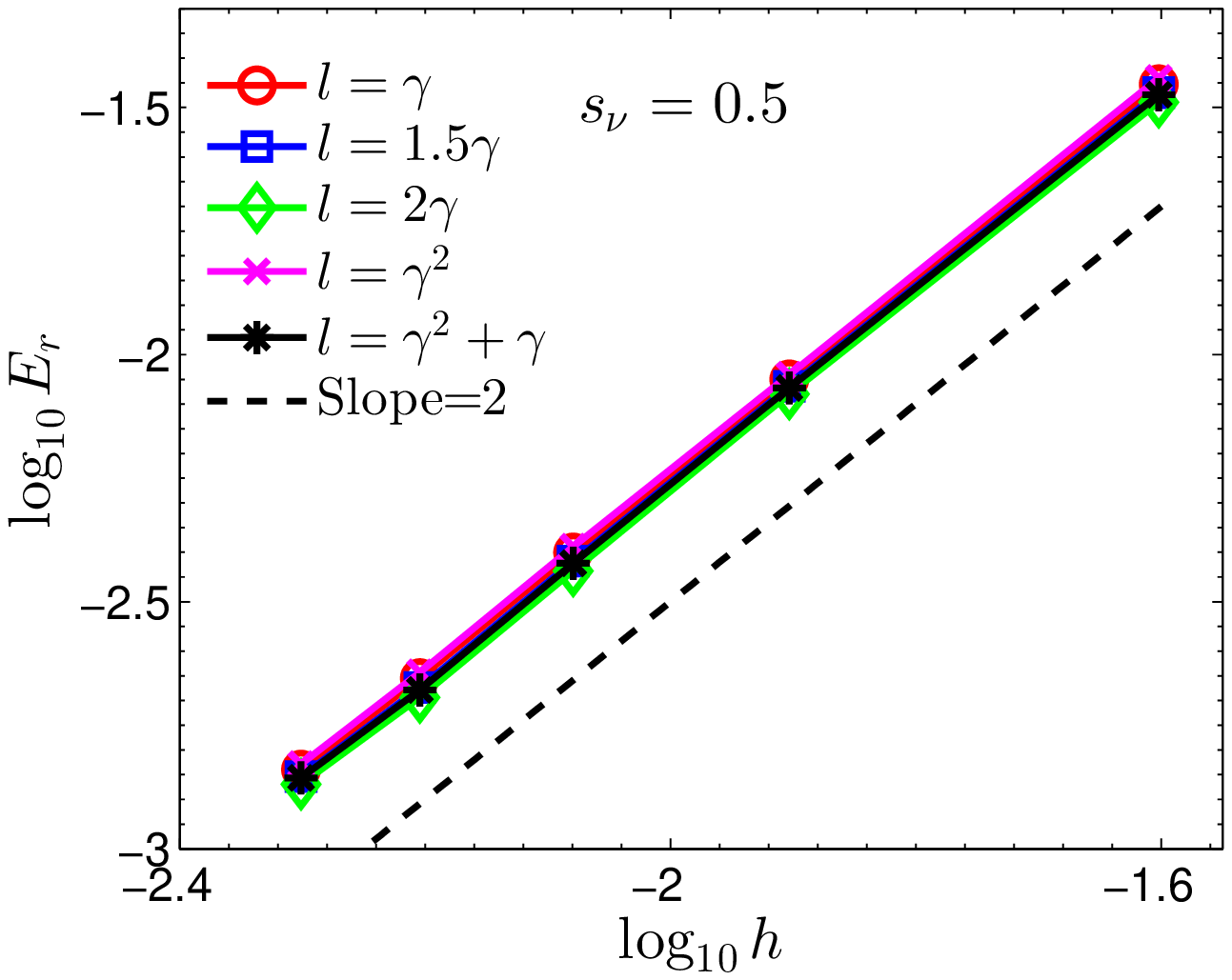}
\includegraphics[scale=0.4]{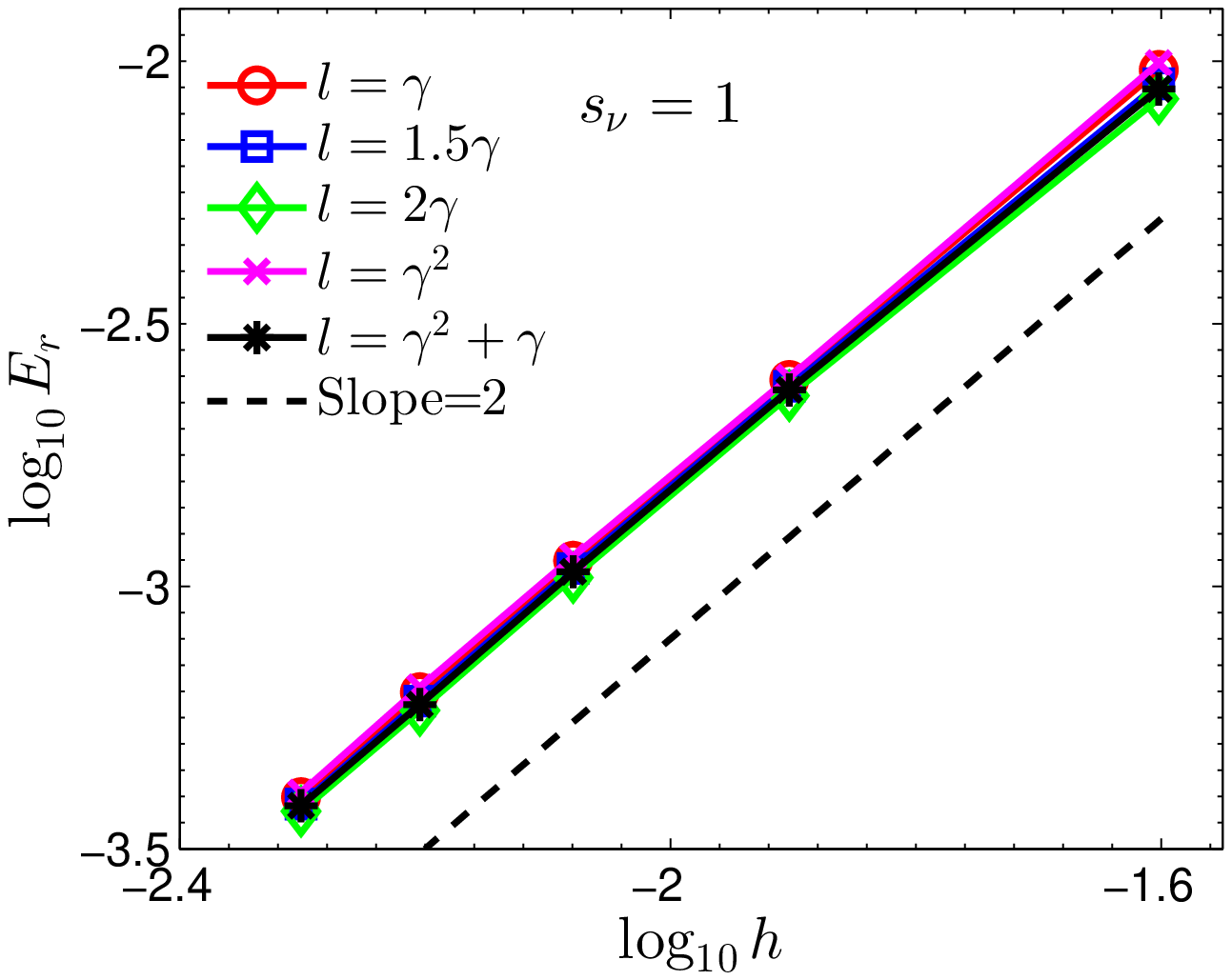}
\includegraphics[scale=0.4]{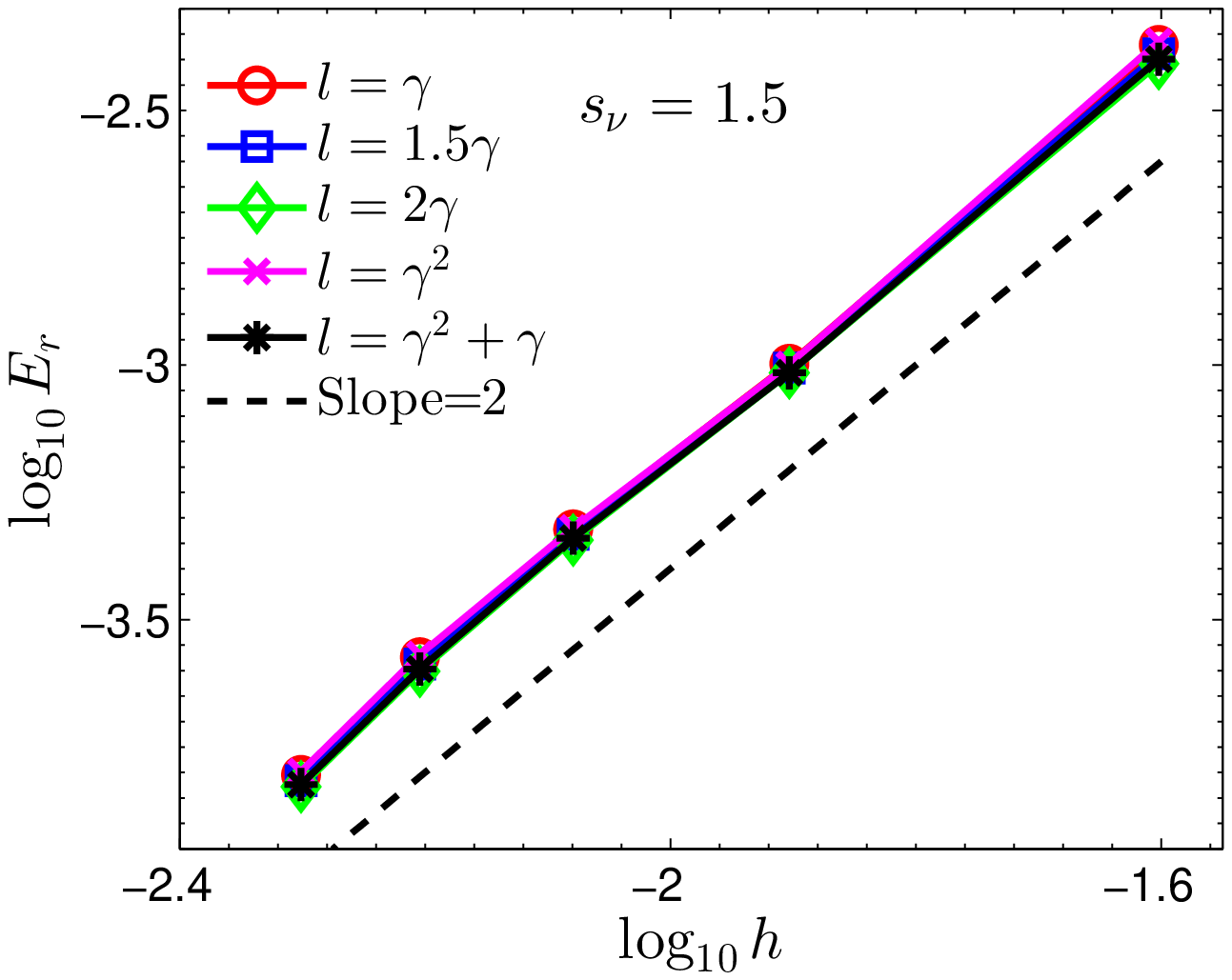}
\includegraphics[scale=0.4]{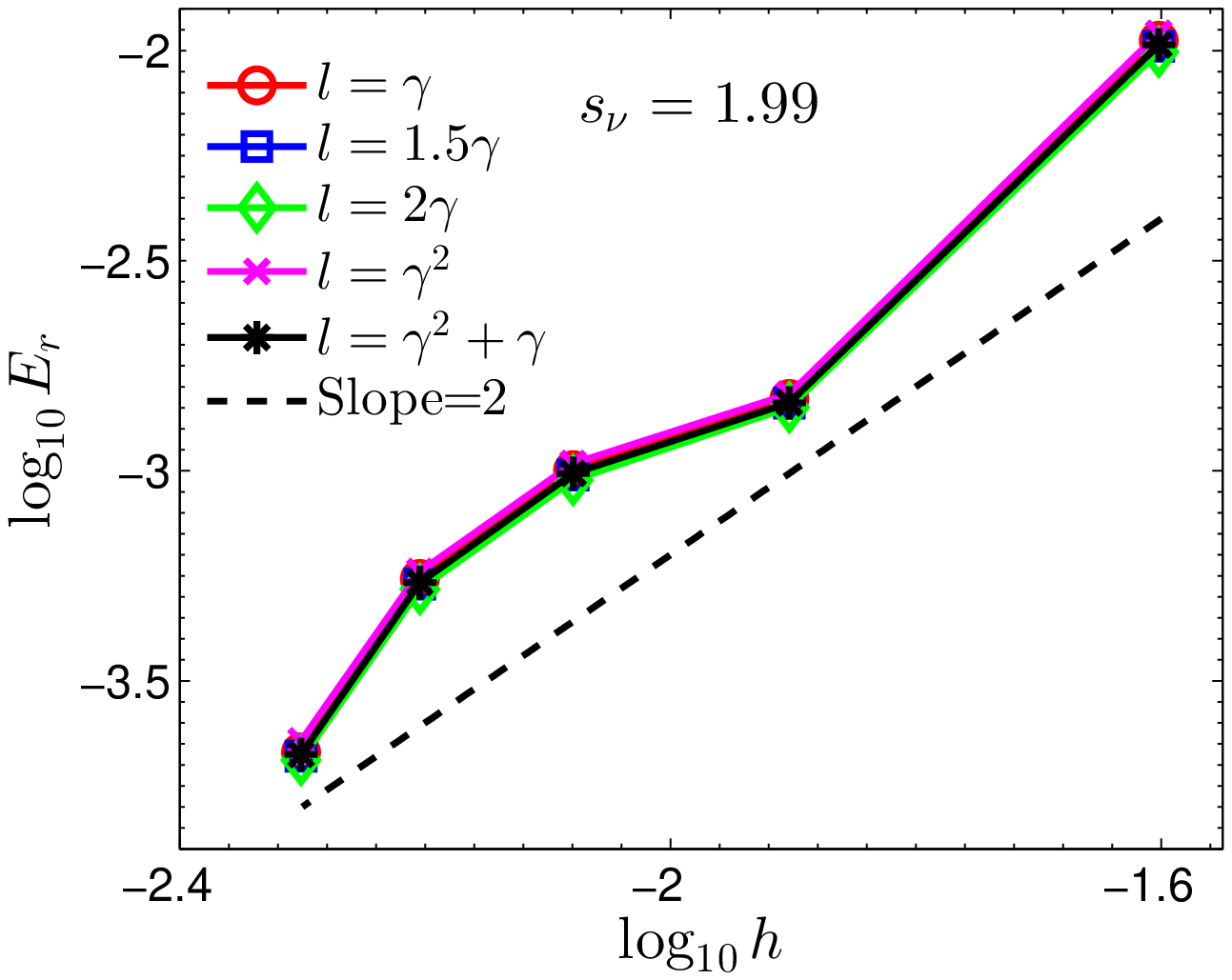}
\caption{Convergence order of the boundary schemes for the Taylor-Green vortex flow. }
\label{Fig:ConvergenOfTaylorGreen}
\end{center}
\end{figure}

To examine the stability and accuracy of the boundary schemes, we take different $s_{\nu}$ ($= 0.5,1,1.5,1.99$) in the simulation with a number of spatial steps $h=1/40$, $1/80$, $1/120$, $1/160$ and $1/200$.
Fig.~\ref{Fig:ConvergenOfTaylorGreen} shows that even with the curved boundary $\partial\Omega$, all the five schemes have second-order accuracy with different $s_{\nu}$. These and the results of the Poiseuille flow show the second-order accuracy and good stability of the convex scheme \eqref{210} for the 2D MRT models.

\subsection{3D Hagen-Poiseuille flow}

For the third problem, we intend to test the schemes for the 3D MRT model.
To this end, we consider the 3D Hagen-Poiseuille flow through a pipe of uniform circular cross-section,
which is an extension of the Poiseuille flow in 2D.
In this situation, there is an external force $\bm F = G(1, 0, 0)$ along the axial direction ($x$-direction) of the pipe.
The problem has the following analytical solution ($\bm u=(u, v, w)$)
\begin{equation}
        u = u(r) = U  ( 1-\frac{r^2}{R^2} )  ,  \quad  v = 0, \quad w =0,
\end{equation}
where $r\in[0, R]$ is the distance to the center line, $R$ is the radius of the circular cross-section
and $U = G R^2 / 4 \nu$ is the maximal velocity along the center line of the pipe.
In the simulation, we take
\begin{equation*}
          \nu=0.03, \quad  G = 0.8\nu, \quad R = \frac{1}{2}.
\end{equation*}

\begin{figure}[!ht]
\begin{center}
\includegraphics[scale=0.4]{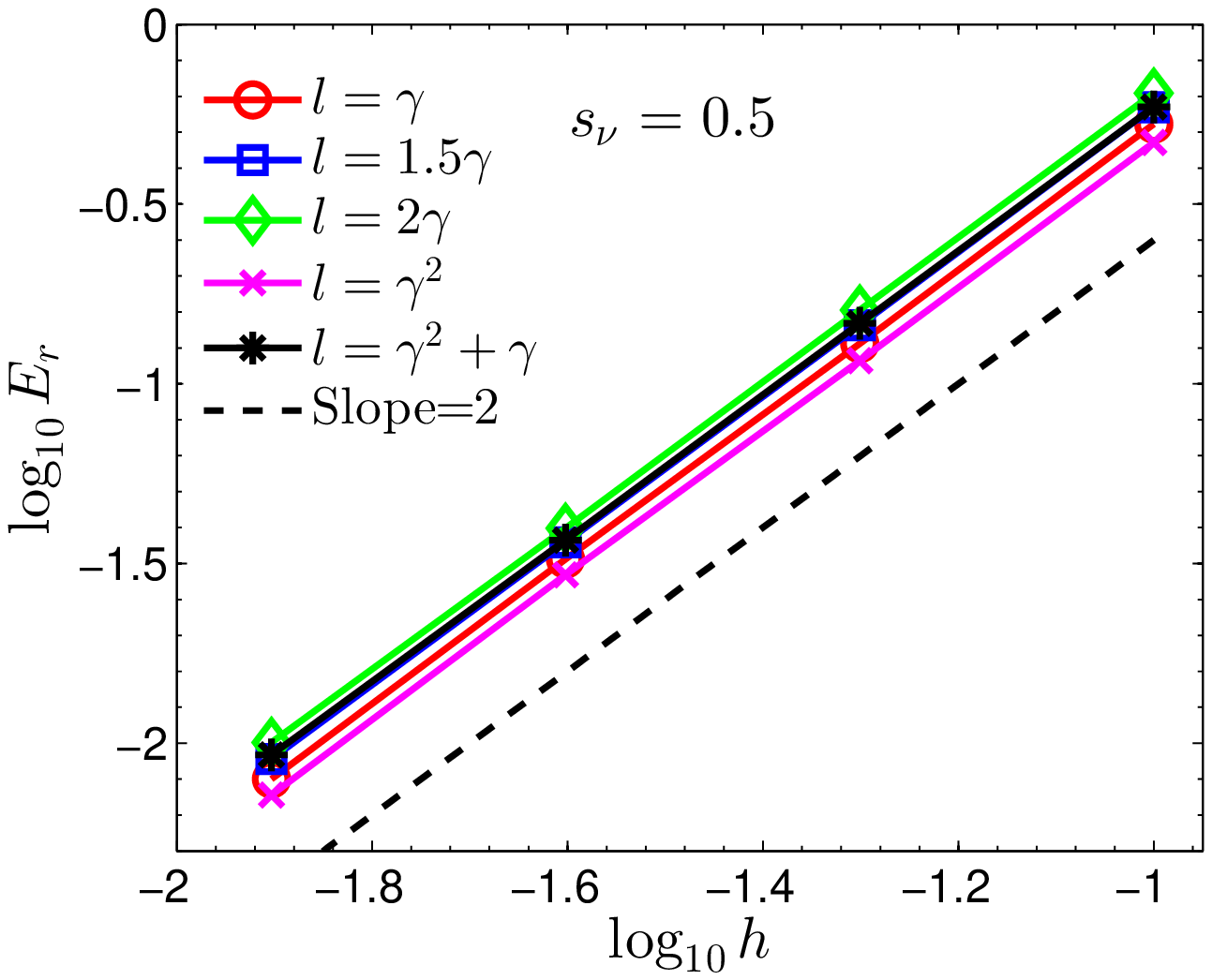}
\includegraphics[scale=0.4]{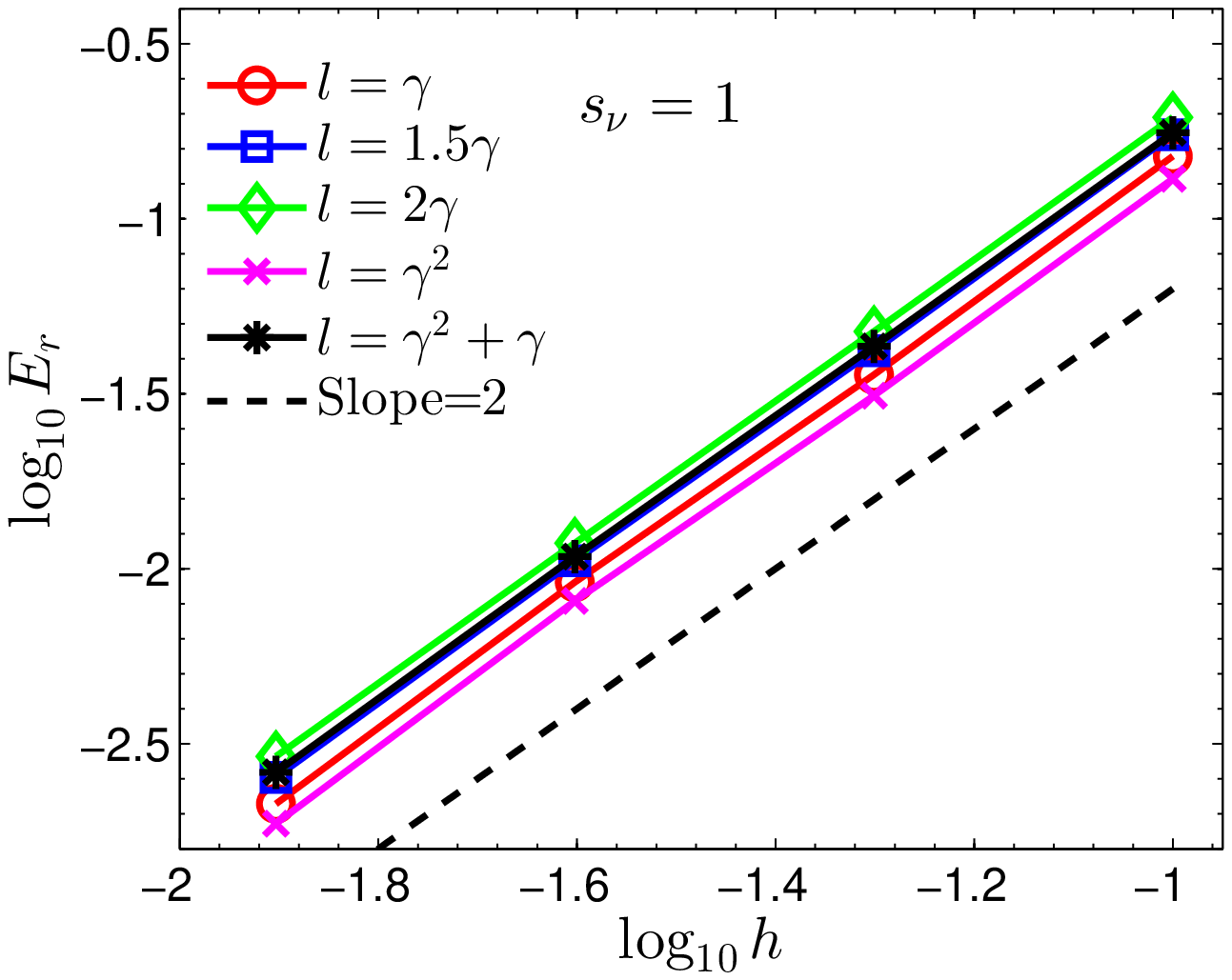}
\includegraphics[scale=0.4]{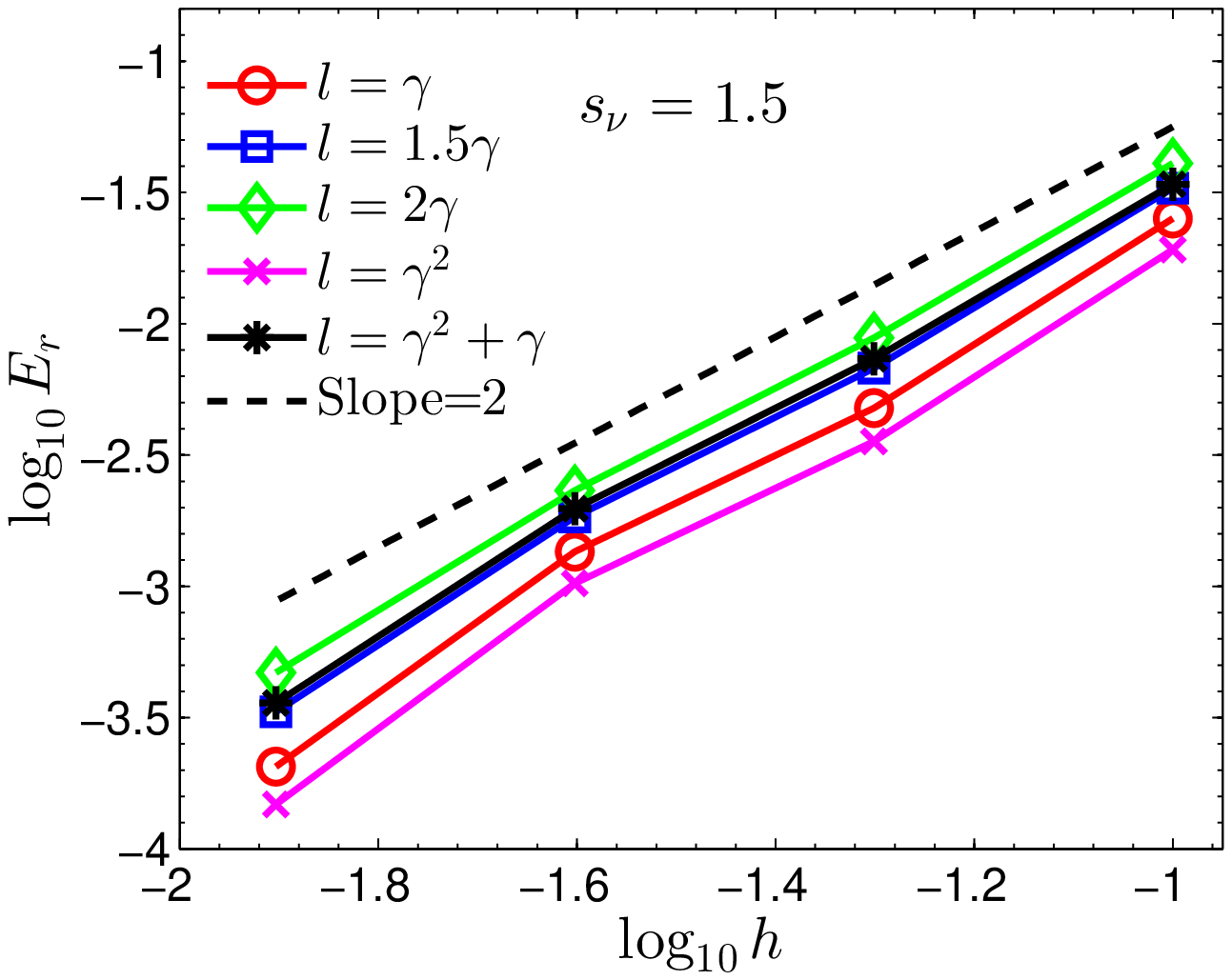}
\includegraphics[scale=0.4]{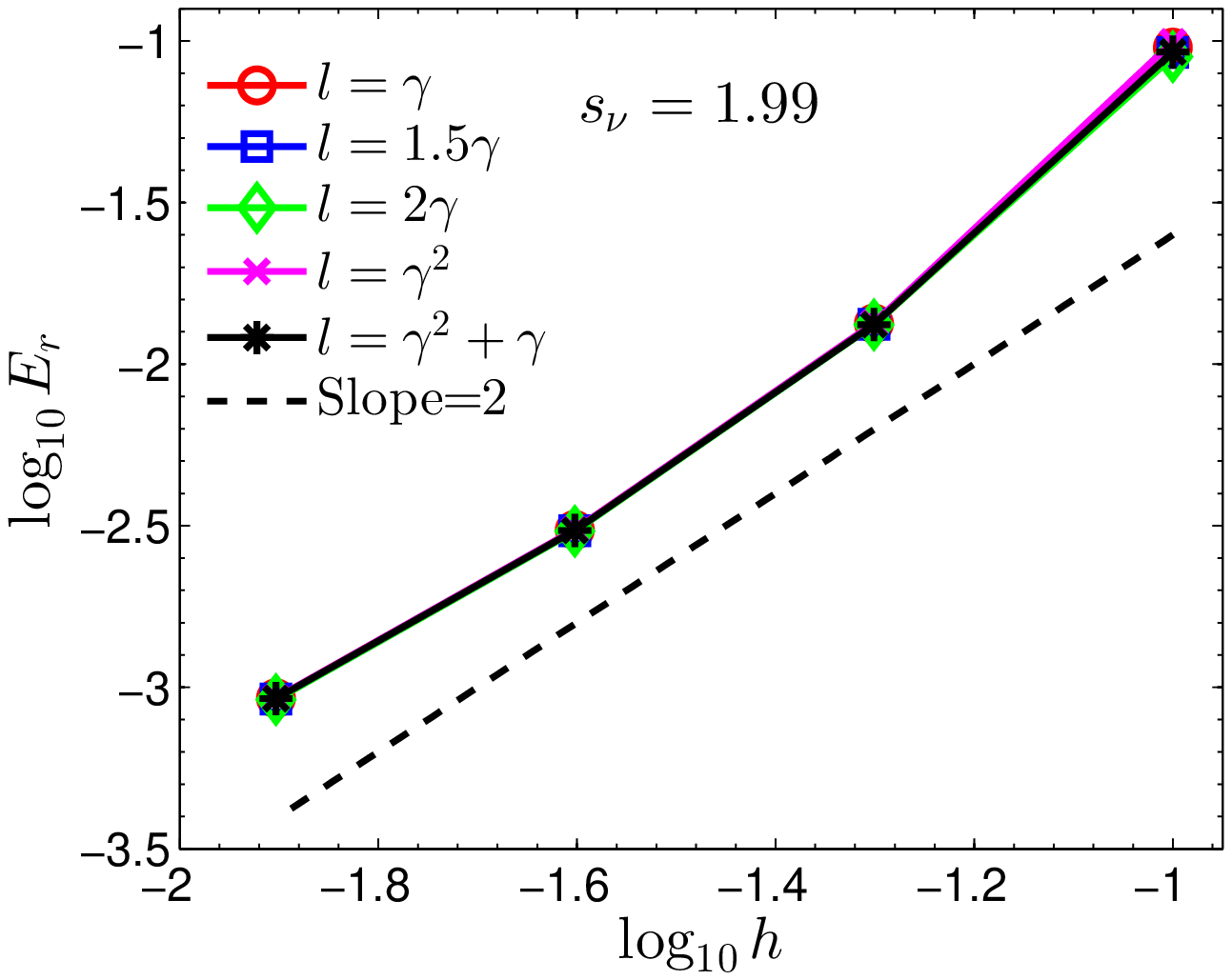}
\caption{Convergence order of the boundary schemes for the 3D Hagen-Poiseuille flow. }
\label{Fig:ConvergenOfHagenPoiseuille}
\end{center}
\end{figure}

Like that for the Poiseuille flow, the axial direction is periodic and
the boundary schemes are applied at the wall of the pipe.
We take different $s_{\nu}$ (=0.5,1,1.5,1.99) and $ h = 1/10,1/20,1/40,1/80$, and the error is computed as in Eq.~\eqref{eqn:L2-error}.
The numerical results are given in Fig.~\ref{Fig:ConvergenOfHagenPoiseuille}. It can be seen that all the five schemes are stable and have second-order accuracy for different $s_{\nu}$. Thus, the good stability and accuracy of the boundary scheme \eqref{210} for the 3D MRT model are validated.

\section{Conclusions and remarks}
\label{sec4}

In this work, we propose a family of single-node second-order boundary schemes for the LBM with general collision models.
The schemes are constructed by generalizing the idea from \cite{GSPK2015,YMS2003} and use the half-way bounce-back scheme as a central step.
The constructed schemes are all second-order accurate for both straight and curved boundaries if so is the bounce-back rule.
the proposed schemes have second-order accuracy for both straight and curved boundaries.
In addition, the schemes are all convex combinations of distribution functions and thereby have good stability.
Finally, numerical experiments are conducted to validate the second-order accuracy and stability of five specific schemes for both 2D and 3D MRT models .

We would like to point out that our schemes contain the existing single-node schemes in \cite{ZY2017,GSPK2015} as special cases but significantly differ from them.
Unlike those for specific TRT models \cite{ZY2017}, the construction of the present schemes are quite universal and simple, it does not involve concrete lattice Boltzmann models.
Our new schemes are also different from that proposed in \cite{GSPK2015} where the half-way bounce-back rule is used only at the boundary point.


\section*{Appendix}

In this appendix, we list the details of the D2Q9 and D3Q15 MRT collision models used in the computations.
The MRT model has the following general form
 $$
 \Omega_i (\bm{x}, \,  t ) =  - \sum_{j} \left(  \mathsf{M}^{-1} \mathsf{S} \mathsf{M} \right)_{ij}  ( f_j - f_j^{(eq)} ) (\bm{x},\, t),
 $$
where $\mathsf{M} \in \mathbb{R}^{q\times q}$ is the transformation matrix, $\mathsf{S} = \diag(s_0,s_1, \ldots, s_{q-1})$ is the diagonal relaxation matrix and
 $f_i^{(eq)}:=f_i^{(eq)}(\bm{x},\, t)$ is the equilibrium given by \cite{He1997jsp}
 \begin{equation}\label{42}
          f_{i}^{(eq)}
          =\omega_i
           \left\{
                  \rho
                  +
                  \rho_0 \left[
                          \frac{\bm c_i \cdot \bm u}{c_{s}^{2}}
                          +\frac{\left(\bm c_i \cdot \bm u \right)^{2}}{2c_{s}^{4}}-\frac{u^{2}}{2c_{s}^{2}}
                  \right]
          \right\}.
\end{equation}
Here $\{\omega_i\}$ are the weight coefficients; $\rho_0$ is the mean density;
$c_s = c / \beta$ is the sound speed with $c:=h / \delta_t$ and $\beta$ a positive const; $\bm c_i = c \bm e_i$,
 $\rho$ and $\bm{u}$ are the fluid density and velocity defined by
\begin{equation*}
  \rho
  =
  \sum_i f_i
  ,
  \qquad
  \rho_0 \bm{u}
  =
  \sum_i \bm{c}_{i} f_i
  .
\end{equation*}

For the D2Q9 model, the discrete velocities are
\begin{equation*}
\left(
\begin{aligned}
e_{ix} \\
e_{iy}
\end{aligned}
\right)
=
\left(
\begin{aligned}
0 && 1 && 0 && -1 && 0 && 1 && -1&& -1&& 1 \\
0 && 0 && 1 && 0  && -1&& 1 && 1 && -1&& -1
\end{aligned}
\right),
\end{equation*}
where $e_{i\alpha}$ is the component of $\bm e_i$ along $\alpha$-direction.
The weight coefficients are
$\omega_0 = 4/9$, $\omega_{1,2,3,4} = 1/9$ and $\omega_{5,6,7,8} = 1/36$ and the sound speed is $c_s=c/\sqrt{3}$.
The transformation matrix is given by \cite{LL2000}:
\begin{equation}\label{eqn:matrix-m}
  \mathsf{M}
  =
  \left(
  \begin{array}{*{9}{r}}
    1 & 1 & 1 & 1 & 1 & 1 & 1 & 1 & 1
    \\
    -4 & -1 & -1 & -1 & -1 & 2 & 2 & 2 & 2
    \\
    4 & -2 & -2 & -2 & -2 & 1 & 1 & 1 & 1
    \\
    0 & 1 & 0 & -1 & 0 & 1 & -1 & -1 & 1
    \\
    0 & -2 & 0 & 2 & 0 & 1 & -1 & -1 & 1
    \\
    0 & 0 & 1 & 0 & -1 & 1 & 1 & -1 & -1
    \\
    0 & 0 & -2 & 0 & 2 & 1 & 1 & -1 & -1
    \\
    0 & 1 & -1 & 1 & -1 & 0 & 0 & 0 & 0
    \\
    0 & 0 & 0 & 0 & 0 & 1 & -1 & 1 & -1
  \end{array}
  \right)
  .
\end{equation}

The discrete velocities for the D3Q15 MRT model are
\begin{equation*}
\left(
\begin{aligned}
e_{ix} \\
e_{iy} \\
e_{iz}
\end{aligned}
\right)
=
\left(
\begin{aligned}
0 && 1 && -1 && 0 && 0 && 0 && 0&&  1&& -1 && 1  && -1 && 1  && -1 && 1 && -1 \\
0 && 0 && 0  && 1 && -1&& 0 && 0&&  1&& 1  && -1 && -1 && 1  && 1  && -1&& -1 \\
0 && 0 && 0  && 0 && 0 && 1 && -1&& 1&& 1  && 1  && 1  && -1 && -1 && -1&& -1
\end{aligned}
\right),
\end{equation*}
the weight coefficients are
$\omega_0 = 2/9$, $\omega_{1-6} = 1/9$ and $\omega_{7-14} = 1/72$ and the sound speed is $c_s=c/\sqrt{3}$.
The transformation matrix corresponding to the above order of discrete velocities is \cite{DGKLL2002}
\begin{equation}\label{eqn:matrix-m3}
  \mathsf{M}
  =
  \left(
  \begin{array}{*{15}{r}}
               1  &  1  &  1  &  1  &  1  &  1  &  1  &  1  &  1  &  1  &  1  &  1  &  1  &  1  &  1 \\
               -2 &  -1 &  -1 & -1  &  -1 &  -1 & -1  &  1  &  1  &  1  &  1  &  1  &  1  &  1  &  1 \\
               16 &  -4 &  -4 & -4  &  -4 &  -4 & -4  &  1  &  1  &  1  &  1  &  1  &  1  &  1  &  1 \\
               0  &  1  &  -1 &  0  &  0  &  0  &  0  &  1  &  -1 &  1  &  -1 &  1  & -1  &  1  & -1 \\
               0  &  -4 &  4  &  0  &  0  &  0  &  0  &  1  &  -1 &  1  &  -1 &  1  & -1  &  1  & -1 \\
               0  &  0  &  0  &  1  &  -1 &  0  &  0  &  1  &  1  &  -1 &  -1 &  1  &  1  &  -1 & -1 \\
               0  &  0  &  0  &  -4 &  4  &  0  &  0  &  1  &  1  &  -1 &  -1 &  1  &  1  &  -1 & -1 \\
               0  &  0  &  0  &  0  &  0  &  1  &  -1 &  1  &  1  &  1  &  1  &  -1 &  -1 &  -1 & -1 \\
			   0  &  0  &  0  &  0  &  0  &  -4 &  4  &  1  &  1  &  1  &  1  &  -1 &  -1 &  -1 & -1 \\
			   0  &  2  &  2  &  -1 &  -1 &  -1 &  -1 &  0  &  0  &  0  &  0  &  0  &  0  &  0  &  0 \\
			   0  &  0  &  0  &  1  &  1  &  -1 &  -1 &  0  &  0  &  0  &  0  &  0  &  0  &  0  &  0 \\
			   0  &  0  &  0  &  0  &  0  &  0  &  0  &  1  &  -1 &  -1 &  1  &  1  &  -1 &  -1 &  1 \\
			   0  &  0  &  0  &  0  &  0  &  0  &  0  &  1  &  1  &  -1 &  -1 & -1  &  -1 &  1  &  1 \\
			   0  &  0  &  0  &  0  &  0  &  0  &  0  &  1  &  -1 &  1  &  -1 &  -1 &  1  &  -1 &  1 \\
			   0  &  0  &  0  &  0  &  0  &  0  &  0  &  1  &  -1 & -1  &   1 &  -1 &  1  &  1  &  -1
  \end{array}
  \right)
  .
\end{equation}

\section*{Acknowledgements}
\label{sec:Conclusions}

The second author (W.-A. Yong) was financially supported
by the National Natural Science Foundation of China (NSFC 11471185) and by the Tsinghua University Initiative Scientific
Research Program (20151080424).



\begin{thebibliography}{99}


\bibitem{LKS}
L.-S.~Luo, M.~Krafczyk, W.~Shyy, in: Encyclopedia
of Aerospace Engineering, edited by R.~Blockley and W.~Shyy, Wiley,
New York, 2010, Chap.~56, 651--660.


\bibitem{YMLS}
D.~Yu, R.~Mei, L.-S.~Luo, W.~Shyy,
\newblock Vicous flow computations with the method of lattice Boltzmann equation,
\newblock { Prog. Aerospace Sci.} 39 (5) (2003) 329--367.

\bibitem{CD}
S.~Chen, G.~D.~Doolen,
\newblock Lattice Boltzmann method for fluid flows,
\newblock { Ann. Rev. Fluid Mech.} 30 (1) (1998) 329--364.


\bibitem{HL1997-1}
X.~He, L.-S.~Luo,
\newblock A priori derivation of the lattice Boltzmann equation,
\newblock  Phys. Rev. E 55(6) (1997) R6333.


\bibitem{HL1997-2}
X.~He, L.-S.~Luo,
\newblock Theory of the lattice Boltzmann method: From the Boltzmann equation to the lattice Boltzmann equation,
\newblock Phys. Rev. E 56(6) (1997) 6811.


\bibitem{Ladd1}
A.~J.~C.~Ladd,
\newblock Numerical simulatons of particulate suspensions via a discretized Boltzmann equation. Part 1. Theoretical Foundation,
\newblock J. Fluid Mech. 271 (1994) 285--309.

\bibitem{Ladd2}
A.~J.~C.~Ladd,
\newblock Numerical simulatons of particulate suspensions via a discretized Boltzmann equation. Part 2. Numerical results,
\newblock J. Fluid Mech. 271 (1994) 311--339.


\bibitem{Jacqmin1999}
D.~Jacqmin,
\newblock Calculation of two-phase Navier-Stokes flows using phase-field modeling,
\newblock J. Comput. Phys. 155 (1999) 96--127.


\bibitem{Huang2013}
J.-J. Huang, H. Huang and X. Wang,
\newblock Wetting boundary conditions in phase-field-based simulation of binary fluids: some comparative studies and new development,
\newblock Int. J. Numer. Meth. Fluids 77 (2014) 123--158.


\bibitem{Bogner2015}
S.~Bogner, R.~Ammer, U.~R\"ude,
\newblock Boundary conditions for free interfaces with the lattice Boltzmann method,
\newblock J. Comput. Phys. 297 (2015) 1--12.

\bibitem{Junk2005pre}
M.~Junk, Z.~Yang,
\newblock One-point boundary condition for the lattice Boltzmann method,
\newblock Phys. Rev. E  72 (6) (2005) 066701.

\bibitem{Noble1995}
D.~R.~Noble, S.~Chen, J.~G.~Georgiadis, R.~O.~Buckius,
\newblock A consistent hydrodynamic boundary condition for the lattice Boltzmann method
\newblock Phys. Fluids 7 (7) (1995) 203--209.

\bibitem{Inamuro1995}
T.~Inamuro, M.~Yoshino, F.~Ogino,
\newblock A non-slip boundary condition for lattice Boltzmann simulations,
\newblock Phys. Fluids 7 (12) (1995) 2928--2930.


\bibitem{Ginzeburg1996}
I.~Ginzburg, D.~d'Humi\`eres,
\newblock Local second-order boundary method for lattice Boltzmann models,
\newblock { J. Stat. Phys.} 84 (5) (1996) 927--971.

%
%
%
%
%
%
%
%


\bibitem{GSPK2015}
M.~Geier, M.~Sch\"onherr, A.~Pasquali, M.~Krafczky,
\newblock The cumulant lattice Boltzmann equation in three dimensiond: Theory and validation,
\newblock Comput. Math. Appl. 70 (2015) 507--547.


\bibitem{ZY2017}
W.~Zhao, W.-A.~Yong,
\newblock Single-node second-order boundary schemes for the lattice Boltzmann method,
\newblock J. Comput. Phys. 329 (2017) 1--15.

\bibitem{Yong2016pre}
W.-A.~Yong, W.~Zhao, L.-S.~Luo,
\newblock Theory of the lattice Boltzmann method: Derivation of macroscopic equations via the Maxwell iteration,
\newblock Phys. Rev. E  93 (2016) 033310.




\bibitem{IGinzburg2005}
I.~Ginzburg,
\newblock Equilibrium-type and link-type lattice Boltzmann models for generic advection and anisotropic-dispersion equation,
\newblock Adv. Water Res. 28 (11) (2005) 1171--1195.

\bibitem{IGinzburg2008_1}
I.~Ginzburg, F.~Verhaeghe, D.~d'Humi\`eres,
\newblock Two-relaxation-time lattice Boltzmann scheme: about parametetrization, velocity, pressure and mixed boundary conditions,
\newblock Commun. Comput. Phys. 3 (2008) 427--478.

\bibitem{IGinzburg2008_2}
I.~Ginzburg, F.~Verhaeghe, D.~d'Humi\`eres,
\newblock Study of simple hydrodynamic solutions with the two-relation-times lattice Boltzmann scheme,
\newblock Commun. Comput. Phys. 3 (2008)  519--581.

\bibitem{YMS2003}
D.~Yu, R.~Mei and W.~Shyy,
\newblock A unified boundary treament in lattice Boltzmann method,
\newblock AIAA Paper, 2003-0953 (2003).

\bibitem{ZY}
W.~Zhao, W.-A.~Yong,
\newblock On the second-order accuracy of the half-way bounce-back rule for the lattice Boltzmann method,
\newblock in preparation.


\bibitem{dHumieres1992rgd}
D. d'Humi\`eres, in Rarefied Gas Dynamics: Theory and Simulations, Prog. Astronaut. Aeronaut.,
Vol. 159, edited by B. D. Shizgal and D. P. Weave (AIAA, Washington, D.C., 1992)
p. 450.

\bibitem{LL2000}
P.~Lallemand, L.-S.~Luo,
\newblock Theory of the lattice Boltzmann method: Dispertion, dispation, isotropy, Galilean invariance, and stability,
\newblock Phys. Rev. E 61 (2000) 6546--6562.

\bibitem{DGKLL2002}
D.~d'Humi\`eres, I.~Ginzburg, M.~Krafczky, P.~Lallemand, L.-S.~Luo,
\newblock Multiple-relaxation-time lattice Boltzmann models in three dimensions,
\newblock Phil. Trans. R. Soc. Lond. A 360 (2002) 437--451.

%
%
%
%
%


\bibitem{Chai2016}
Z.~Chai, C.~Huang, B.~Shi, Z.~Guo,
\newblock A comparative study on the lattice Boltzmann models for predicting effective diffusivity of porous media
\newblock Int. J. Heat Mass Tran. 98 (2016) 687--696.

\bibitem{Fakhari2017}
A.~Fakhari, D.~Bolster,
\newblock Diffuse interface modeling of three-phase contact line dynamics on curved boundaries: A lattice Boltzmann model for large density and viscosity ratios.
\newblock J. Comput. Phys. 334 (2017) 620--638.


\bibitem{He1997jsp}
X.~He, L.-S.~Luo,
\newblock Lattice Boltzmann model for the incompressible Navier-Stokes equation,
\newblock J. Stat. Phys. 88 (3) (1997) 927--944.







\end{thebibliography}
\end{document}